\documentclass[a4paper,fleqn,usenatbib]{mnras}
\pdfoutput=1
\usepackage{txfonts}

\usepackage[T1]{fontenc}
\usepackage{ae,aecompl}
\usepackage{longtable}

\usepackage{graphicx}	
\usepackage{amssymb}	
\usepackage{pdflscape}
\usepackage{epstopdf}
\usepackage{ulem}

\usepackage{etoolbox}
\makeatletter
\patchcmd\@combinedblfloats{\box\@outputbox}{\unvbox\@outputbox}{}{%
	\errmessage{\noexpand\@combinedblfloats could not be patched}%
}%
\makeatother
\newcommand{\flux}{\,ergs \,cm$^{-2}$ \,s$^{-1}$ \,\AA $^{-1}$}
\newcommand{\angstrom}{\AA}

\title[UV stellar population of the old open cluster M67]{ULTRAVIOLET STELLAR POPULATION OF THE OLD OPEN CLUSTER M67 (NGC 2682)}

\author[Sindhu N et al.]{
Sindhu. N,$^{1,}$$^{2}$\thanks{E-mail:sindhu.n@iiap.res.in}
Annapurni. Subramaniam,$^{2}$\thanks{E-mail:purni@iiap.res.in}
and C. Anu Radha$^{1}$
\\
$^{1}$School of Advanced Sciences, VIT University, Vellore, India\\
$^{2}$Indian Institute of Astrophysics, Koramangala, Bangalore, India}

\date{Accepted XXX. Received YYY; in original form ZZZ}

\pubyear{2018}

\begin{document}
\label{firstpage}
\pagerange{\pageref{firstpage}--\pageref{lastpage}}
\maketitle

\begin{abstract}
We present the results of ultraviolet (UV) photometry of the old open cluster M67 obtained using Galaxy Evolution Explorer (GALEX) in far-ultraviolet (FUV) and near-ultraviolet (NUV) bands. UV detections of 18 blue straggler stars (BSSs), 3 white dwarfs (WDs), 4 yellow straggler stars, 2 sub-subgiants, and 25 X-ray sources are presented (along with an online catalog). We demonstrate the capability of UV colour magnitude diagrams (CMDs) along with the UV isochrones to identify potential stars which defy standard stellar evolution in this well studied cluster. We also detect a few main sequence turn-off and subgiant branch stars with excess flux in the FUV and/or NUV. UV continuum excess as well as Mg II {\it h} +{\it k} emission lines from the IUE archival spectra for 2 red giants are detected, suggestive of their chromospheric activity. We suggest that a large number of stars in this cluster are chromospherically active, whereas the bright BSS are unlikely to be active. We also estimate the fundamental parameters L/L$_{\sun}$, R/R$_{\sun}$ and T$_{eff}$ of the BSSs and 15 FUV bright stars by constructing the spectral energy distribution (SED) using multi-wavelength data. We identify three groups among the BSSs, based on their properties. The H-R diagram of BSSs with isochrones suggests that the BSSs in M67 are formed in the last 2.5 Gyr - 400 Myr, more or less continuously. We identify 7 potential MS+WD candidates based on large UV excess from a probable 11, based on SEDs.
\end{abstract}




\section{Introduction}
\label{sec:Intro}
M67 is a benchmark cluster to study stellar evolution, dynamics and cluster properties due to its proximity, richness, solar metallicity and exotic stellar populations. This cluster is located at RA = 8h 51m 23.3s and Dec =11\degr  49\arcmin  02\arcsec\, at a distance of $\approx$ 900 pc and has a reddening of E (B$-$V) = 0.015 to 0.056 mag (\citealp{Janes1984}, \citealp{montgomery93} (henceforth MMJ93), \citealp{Taylor2007}). Estimation of the cluster age varies between 3 to 5 Gyr (MMJ93, \citealp{VandenBerg2004}), with the recent findings indicating it to be 3.5 Gyr (\citealp{chen2014}, \citealp{Bonatto2015}). \citet{Gonzalez2016a} estimated a gyro-age of 3.7 $\pm$ 0.3 Gyr,  \citet{Gonzalez2016b} revised the estimate to 5.4 $\pm$ 0.2 Gyr, whereas \citet{Barnes2016} estimated an age of 4.2 $\pm$ 0.2 Gyr, \citet{Stello2016} estimated a seismic-informed distance of 816 $\pm$ 11 pc, all from the K2 mission data.
The cluster has evolved phases of both single and binary stellar evolution. 38\% of the cluster members are found to be in binaries (MMJ93). The main sequence turn off (MSTO) mass of the cluster is about $\approx$ 1.25M$\sun$ to 1.3M$\sun$ (\citealp{Sandquist2003_EB_TO}, \citealp{Sandquist2004}, \citealp{Gokay2013}). The cluster members are well identified by membership studies using both proper motion (\citealp{Sanders1977}, \citealp{Girard1989}, \citealp{Zhao1993}, \citealp{Yadav2008}, \citealp{Krone-Martins2010}) and radial velocity (\citealp{Matheiu1986}, \citealp{Mathieu1990_SB22}, \citealp{Milone1992}, \citealp{Milone1994}, \citealp{Yadav2008}, \citealp{Pasquini2011}, \citealp{Geller2015}, \citealp{Brucalassi2017}) measurements with various spatial extent and limiting magnitudes. The cluster is exhaustively studied through photometry in the multiwavelength bands (\citealp{Nissen1987}, \citealp{montgomery93}, \citealp{Fan1996}, \citealp{landsman98}, \citealp{Belloni98}, \citealp{VandenBerg2004_Chandra}, \citealp{Sarajedini2009}, \citealp{seigel14}, \citealp{Mooley2015}). 

 Blue straggler stars (BSSs) are cluster members that are brighter, and bluer than stars on the upper main sequence (\citealp{1953Sandage}). The two main leading scenarios proposed for their formation are stellar collisions leading to mergers in high density environments (\citealp{Hills1976}) and mass transfer (MT) from an evolved donor to a lower-mass star in a binary system in low density environments (\citealp{McCrea1964}, \citealp{Chen2008}). An exceptionally large number of 24 BSSs are detected in M67 as compared to any other old open clusters (\citealp{Deng99}). Sub-subgiants (SSGs) are rather a new class of stars (\citealp{Geller2017}), which occupy a unique location in the colour-magnitude diagram (CMD), red ward of the main-sequence (MS) and fainter than the subgiants where normal single star evolution does not predict stars (\citealp{Geller2017_SSG}). Two SSGs (WOCS\footnote[1]{WOCS (WIYN open cluster study) ID is taken from \citet{Geller2015} and we refer to this ID in the entire text, and for stars that were not observed in their study we refer to \citet{montgomery93} with 'MMJ' ID.} 15028 and WOCS 13008)  identified in the cluster are also X-ray sources (\citealp{Belloni98}, \citealp{Mathieu2003}). Yellow straggler stars (YSSs) fall above the subgiant branch in optical CMDs, between the BSSs and the red giants (RG). YSSs may represent a population of evolved BSSs (\citealp{Mathieu1990_SB22}, \citealp{Leiner2016}). Four YSSs (WOCS 2002, WOCS 2008, WOCS 1015 and WOCS 1112) often referred as yellow giants are observed in the cluster CMD (\citealp{Geller2015}). \citet{Landsman1997_S1040} detected one of the YSS (WOCS 2002) with a white dwarf companion, that has undergone a mass transfer, where as \citet{Leiner2016} conclude that a merger or collision is most likely to have occured in WOCS 1015 and thus YSSs are likely to be evolved BSSs. ROSAT, Chandra and XMM - Newton studies of the cluster detected at least 36 member stars with X-ray emission (\citealp{Belloni1993}, \citealp{Belloni98}, \citealp{VandenBerg2004_Chandra}, \citealp{Mooley2015}). Hence, M67 has both single and binary stars, along with some exotic stellar systems. A study of  Ultraviolet (UV) characteristics of these stellar population will help in understanding their formation and evolution.

Previous studies of this cluster in the UV has been done by \citet{landsman98} and \citet{seigel14}. \citet{landsman98} detected 20 stars in M67 which includes 11 BSSs, 7 WD candidates, a YSS$+$WD binary (WOCS 2002) and a non member using Ultraviolet Imaging Telescope (UIT). They also presented a semi-empirical integrated spectrum of M67 showing a domination of BSSs at shorter wavelengths than 2600 \AA\,.  However, UIT images are not deep enough to detect fainter population of stars in the cluster. Recently, \citet{seigel14} studied M67 using the Ultraviolet Optical Telescope (UVOT) on Swift Gamma-Ray Burst Mission. They detected 10 BSSs along with a number of stars near the WD cooling sequence. They showed that UVOT could easily distinguish stellar population such as BSSs, WDs and young \& intermediate age MS stars. 

M67  is well studied in the optical and X-ray, though a deep study in the UV is lacking. The UV observations can detect the presence of possible hot companions that are not evident from optical photometry alone. The presence of a hot component to BSSs was detected by far-UV observations of NGC 188 by the HST (\citealp{gosnell15}). This provided the necessary observational evidence of MT as one of the formation mechanism of BSSs. \citet{2016Subramaniam} identified a post-HB/AGB companion to a BSS in NGC 188 with the help of far-UV and near-UV photometry from the Ultra Violet Imaging telescope (UVIT). In the case of M67, a deep UV study will help in identifying similar systems among the BSSs. The deep far-UV and near-UV data will help in understanding the UV properties of stars, that do not follow the standard single star evolution as well as those of normal stars. 

Galaxy Evolution Explorer (GALEX) observed this cluster in several pointings to produce deep images of this cluster. In this study, we use these images to understand the UV stellar population in M67. We present a comprehensive study of this cluster using FUV and NUV data from GALEX.  

In this study, we present the UV properties of optically detected members of M67 using GALEX photometry. The paper is structured as follows: The UV data and their optical counterparts are described in Section \ref{sec:data}. In section \ref{Analysis} we analyse the UV-Optical CMDs \& UV CMDs and focus on UV bright stars observed in GALEX. In Section \ref{discussion} we present the discussion and summarise with the conclusion in section \ref{Conclusions}.
\section{Data}
\label{sec:data}
\subsection{UV data}
\label{sec:UVdata}
GALEX UV space mission was aimed at both UV imaging and spectroscopic surveys simultaneously in two broad bands, far-UV (FUV; 1344$-$1786 \AA) and near-UV (NUV; 1771$-$2831 \AA). It had a wide field of view of  $\sim$1.3\degr\ and a spatial resolution of 4.2\arcsec\ and 5.3\arcsec\ in FUV and NUV respectively. Further details of its on-orbit performance and satellite are described in \citet{Martin05} \& \citet{Morrissey07}. GALEX performed sky surveys with different depth and coverage; viz. All-sky imaging survey (AIS), Medium-depth imaging survey (MIS), Deep imaging sky survey (DIS) and Guest investigator (GII) survey (\citealp{Morrissey07}, \citealp{Bianchi2009}). 

\begin{figure}
	\includegraphics[width=1.0\columnwidth]{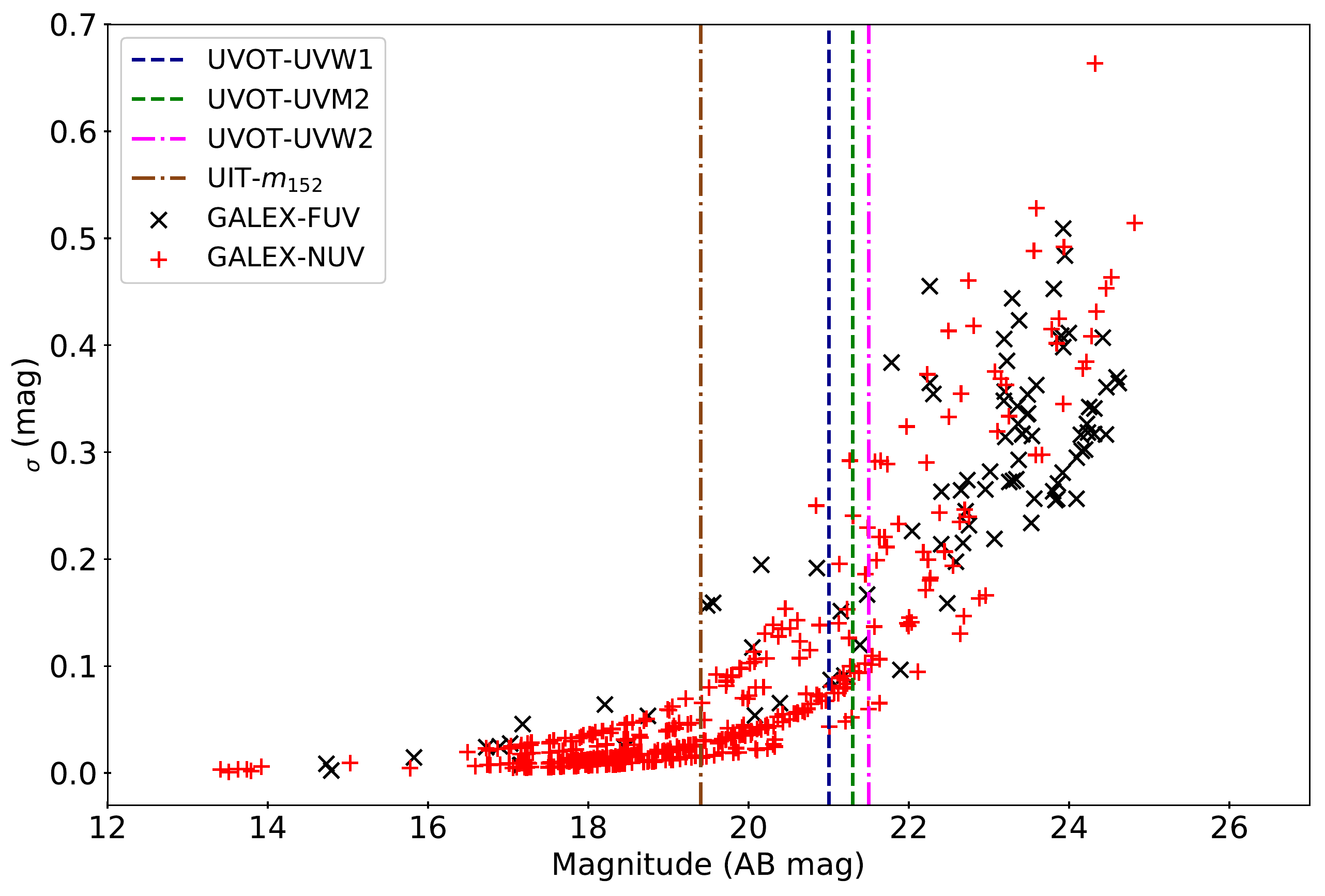}
	\caption{Photometric error in GALEX FUV and NUV data vs. their magnitudes.  The photometric limit of other UV studies of M67 are also indicated.}
	\label{error_plot}
\end{figure}

We obtained the pipeline reduced photometric data of M67 from GR6/GR7 data release. The UV data are mainly obtained from GII and MIS, with a few stars from AIS. Table \ref{tab 1: DATA} lists GALEX observations of M67. We carefully removed data with multiple measurements as well as artefacts, in the data obtained (\citealp{Bianchi14}) from the pipeline. A visual inspection of the images shows that the cluster region is not crowded, and pipeline photometry can be considered reliable. The FUV and NUV magnitudes  are corrected for saturation (\citealp{Camarota2014}). The photometric error in FUV and NUV bands after saturation correction are shown in Figure \ref{error_plot} as a function of magnitude. In this plot, we also show the photometric depth of the UVOT and UIT studies of M67. The photometric depth of UIT (m$_{152}$) is $\sim$ 19.5 mag (\citealp{landsman98}), and those of UVOT are $\sim$ 21.5 mag (\citealp{seigel14}), whereas the GALEX photometry goes down up to $\sim$ 24.5 mag.
\subsection {Optical counterparts}
\label{sec:Opticaldata}
The optical photometric data of M67 are obtained from MMJ93. These stars are cross matched with \citet{Yadav2008} data comprising of stars with a proper motion membership probabilty $\ge 60 \%$ and \citet {Geller2015} data with radial velocity membership probability $\ge 50 \%$. Based on both kinematic and radial velocity membership study, we identified 531 cluster members, such that a star is either a proper motion member and/or a radial velocity member. Among these, 164 stars are members based on proper motion membership probability alone. This is due to either the star not studied by \citet{Geller2015}, as the limiting magnitude is upto 16.5 mag or the star being a probable member/unknown based on radial velocity measurements. Photometry of additional 152 stars are taken from \citet{Geller2015}, that are classified to be members based on radial velocity membership study and are not found in MMJ93. Hence we have selected a total of 683 member stars of the cluster having BV optical photometric data. Their counterparts in the FUV and NUV data are identified with an accuracy of $\le$3\arcsec\ using GalexView\footnote[2]{http://galex.stsci.edu/GalexView/}. A star with the closest angular separation between optical and UV crossmatch is selected in case of multiple detections. Out of 683 cluster members, 92 stars are detected in the FUV and 424 stars in the NUV band. Among the 92 stars detected in FUV, 84 are members based on radial velocity measurement, 4 are based on only proper motion membership probability, 3 are WDs identified by \citet{landsman98} and 1 star (WOCS 1017) with low proper motion and radial velocity membership probabilties. WOCS 1017 is in the location of BSS, it may be interesting to study this object in detail as \citet{landsman98} argues that, as a blue straggler, this star might acquire a peculiar velocity if its formation process is due to stellar interactions (e.g., \citet{Leonard1996}). Of the 424 stars detected in NUV band, 400 are members by radial velocity study, 21 are proper motion members, 2  are WDs identified by \citet{landsman98} and WOCS 1017. The UV photometric data of these stars, along with their already known evolutionary status (classification) taken from literature are tabulated in Table \ref{tab 1:Stars}.
 \begin{table}
	\centering
	\caption{Observation details of M67 by GALEX, which are used in this study.}
	\label{tab 1: DATA}
	\begin{tabular}{ccc} 
		\hline
		No. of stars & Sky survey & Exposure time (s). \\	
		\hline 	
		\multicolumn{3}{c}{FUV band detection}\\
		\hline
		40 & GII & 5555.20\\
		34 & GII & 1691.05 \\
		11 & MIS & 455.00 \\
		7  & AIS & 178.05\\	
		\hline
			\multicolumn{3}{c}{NUV band detection}\\
	    \hline
	    	92 & GII & 5555.20\\
	    	181& GII & 1691.05\\
	    	20 & MIS & 1673.30\\
	    	1 & MIS & 1594.03\\
	    	55 & MIS & 455.00\\
	    	1 & AIS & 455.00\\
	    	5 & AIS & 217.00\\
	    	2 & AIS & 216.00\\
	    	3  & AIS & 208.00\\
	    	1  & AIS & 192.00\\
	    	64 & AIS & 178.05\\
		\hline
	\end{tabular}
\end{table}
\subsection{UV and Optical Isochrones}
\label{sec:FSPS}
\begin{figure*}
	\includegraphics[width=1.3\columnwidth]{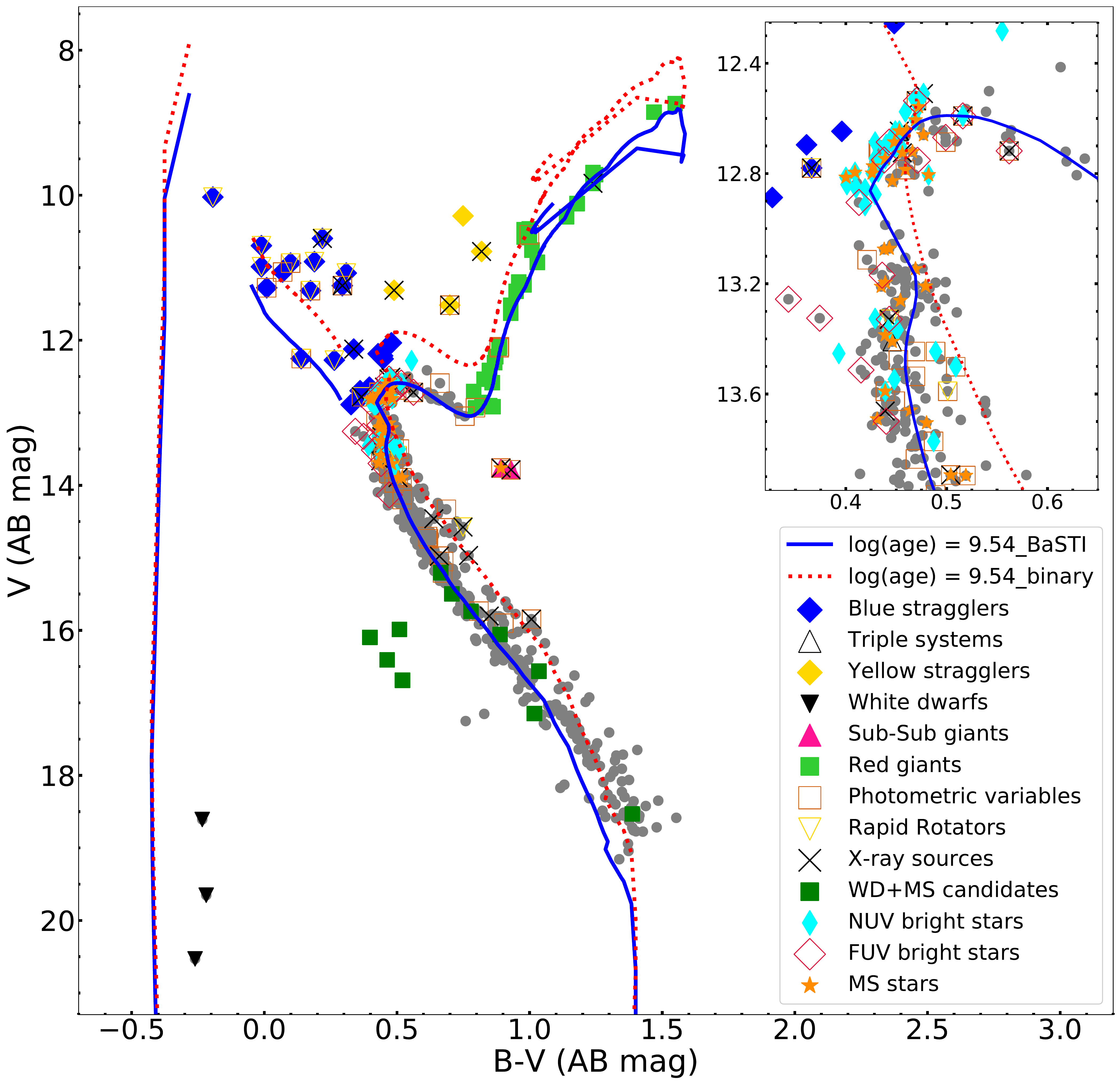}
	\caption{Optical CMD of M67, showing only member stars based on proper motion and radial velocity measurements of the cluster. We use V$-$M$_{v}$= 9.6 mag and E (B$-$V)=0.056. A 3.5 Gyr isochrone is superposed on the optical CMD along with the equal mass binary sequence. The WD cooling sequence is shown to the left of the MS and the BSS model line is shown above the MSTO.}
	\label{optical CMD}
\end{figure*}

We have used flexible stellar population synthesis (FSPS) (\citealp{Conroy2009}, \citealp{Conroy2010}) code to generate both optical and UV isochrones of BaSTI model (A Bag of Stellar Tracks and Isochrones) (\citealp{Pietrinferni2004}, \citealp{Cordier2007}) by providing the input parameters of the cluster viz. distance modulus (V$-$M$_{v}$= 9.6 mag), Solar metallicity and reddening of E (B$-$V)= 0.056 mag. The UV isochrones are obtained after convolving the GALEX filters with the BaSTI model. FSPS code is capable of handling various phases of stellar evolution including the horizontal branch stars, BSSs and thermally pulsing-asymptotic giant branch stars along with the standard evolutionary sequences. Figure \ref{optical CMD} shows the V, (B$-$V) CMD of above mentioned 683 member stars and the stellar population identified based on previous studies. The V and B magnitudes are converted to AB magnitude system as the GALEX magnitudes are in this system, and the transfromation values for vega to AB magnitude system are taken from \citet{Bianchi2011} Table 1. An isochrone of 3.5 Gyr age is superposed on the optical CMD along with a binary sequence, which is 0.75 mag brighter than the MS as shown in Figure \ref{optical CMD}. These isochrones are found to match the observed sequence very well. A WD cooling curve to the left of the MS and BSS model line above the MSTO are shown in the Figure \ref{optical CMD}. The BSS model line is the locus for BSS, assuming them to be MS stars with masses in excess of the turn-off mass, which uniformly populates 0.5 magnitudes above the MSTO to 2.5 magnitudes brighter than the MSTO. The FSPS model generates this model line as well as the WD cooling curve. All the stars classified in Table \ref{tab 1:Stars}, are shown with different coloured symbols in all the CMDs and discussed in the following sections. Three WDs identified by \citet{landsman98} in the UIT images are located near the WD cooling curve. In the optical CMD, 22 BSSs are identified. The brightest BSS (WOCS 1010) lie above the BSS model line, suggesting that it is brighter than the general brightness distribution found among the BSS population. Four YSSs are located in the region between the BSSs and the red giant branch (RGB). Two SSGs are redder than the MS and lie below the subgiant branch in the CMD. We use this optical CMD as the reference CMD to understand the UV CMDs presented in the next section. 

\section{Data Analysis}
 \label{Analysis}
We present the UV$-$optical and UV CMDs of this cluster in this section and discuss the various stellar population that are distinct in these CMDs. We list and discuss the X-ray sources which are detected in GALEX. We also present the UV properties of BSSs in the UV$-$optical and UV CMDs.
\subsection{ UV and UV $-$ optical CMDs}
\label{UV section}
UV$-$optical CMD combines the UV and optical photometry and helps to identify the hot stellar population as well as binaries with hot companions such as WDs.  The number of stars detected in FUV and NUV with optical counterparts are 92 and 424 respectively, whereas only 67 stars have both FUV and NUV detection. In the following section, we analyse the FUV$-$optical CMD, NUV$-$optical CMD and the UV CMDs.
\begin{figure*}
	\centering
\includegraphics[width=1.0\columnwidth]{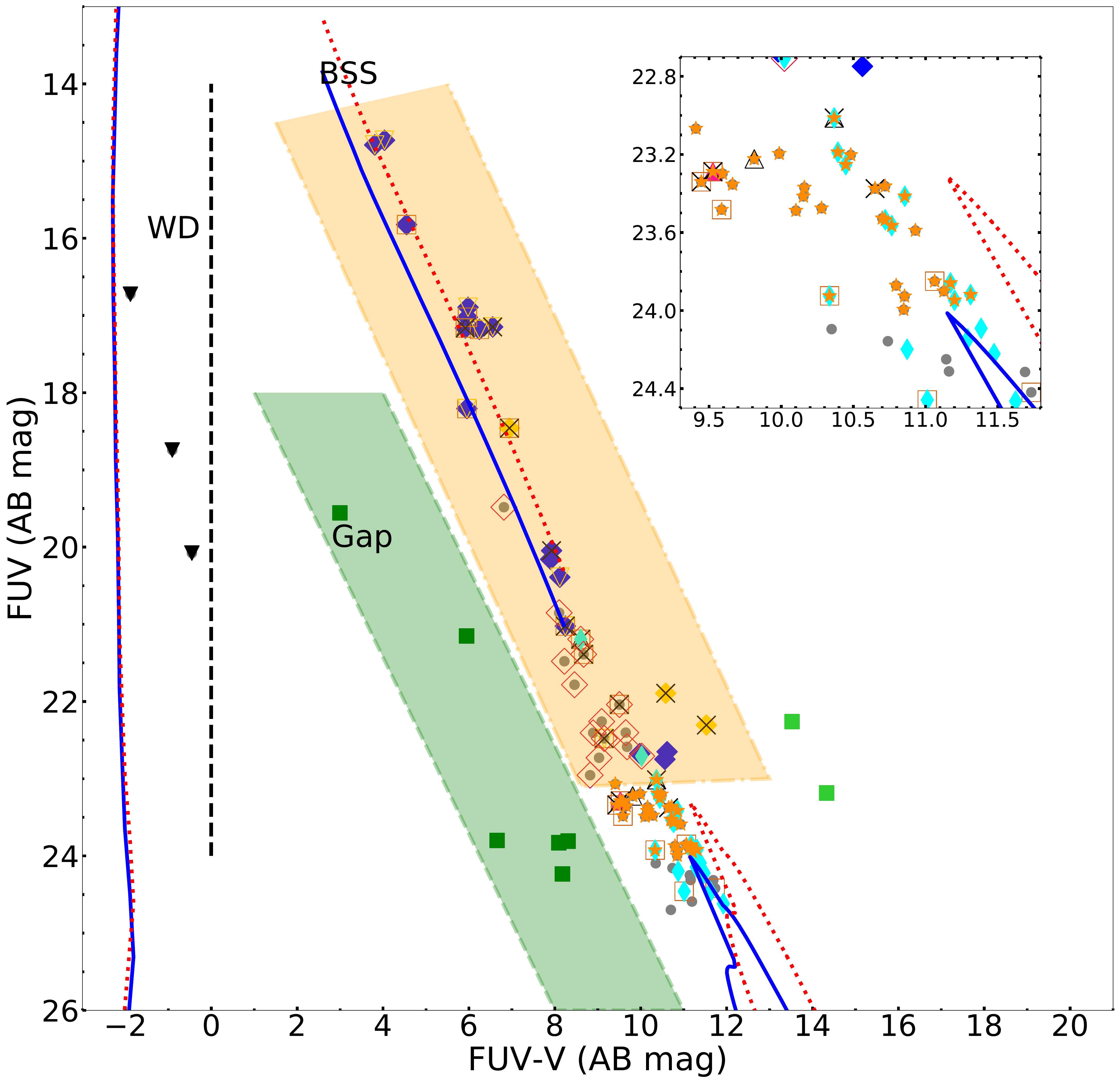}\includegraphics[width= 1.0\columnwidth]{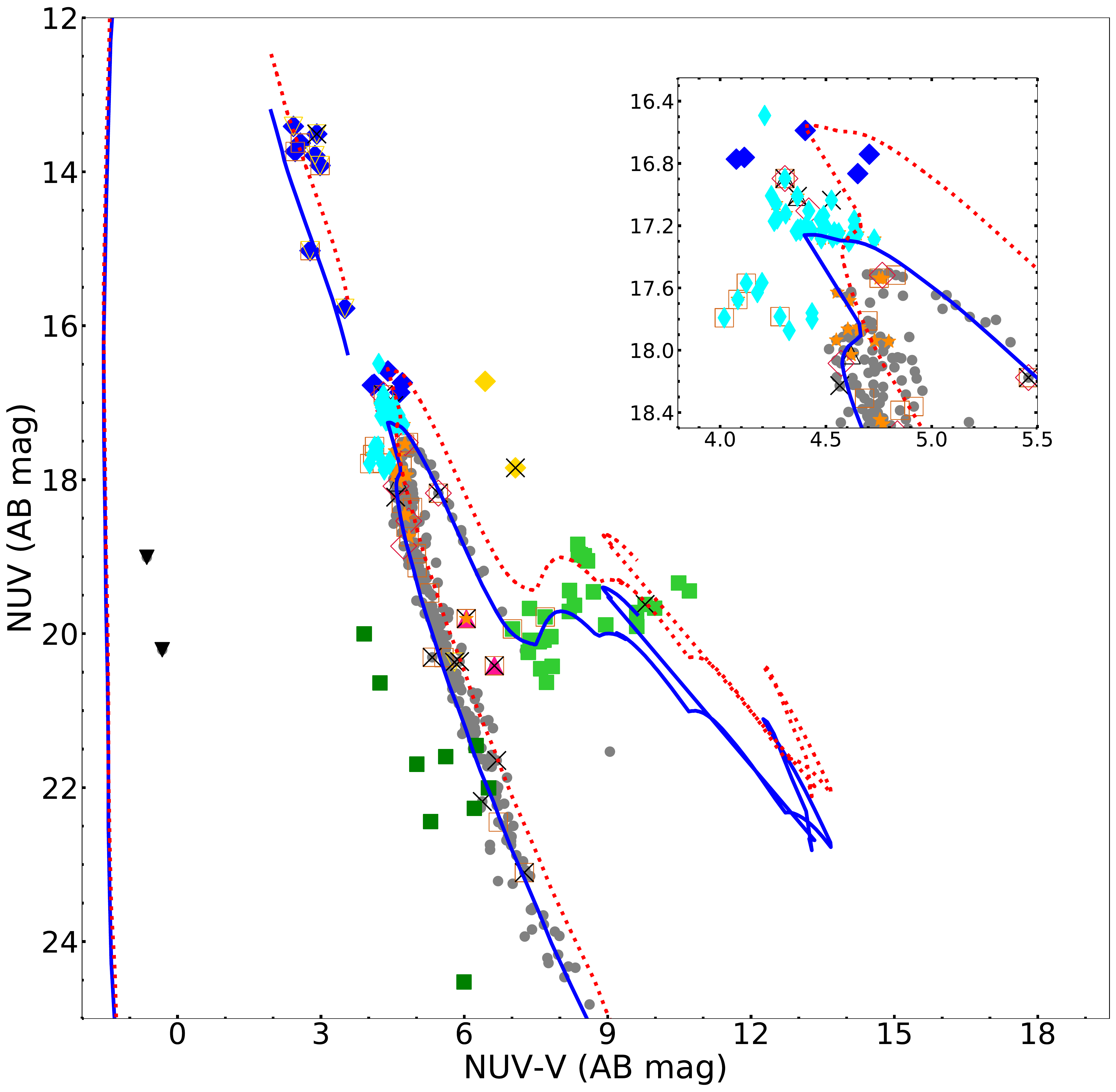}\\
\includegraphics[width= 1.0\columnwidth]{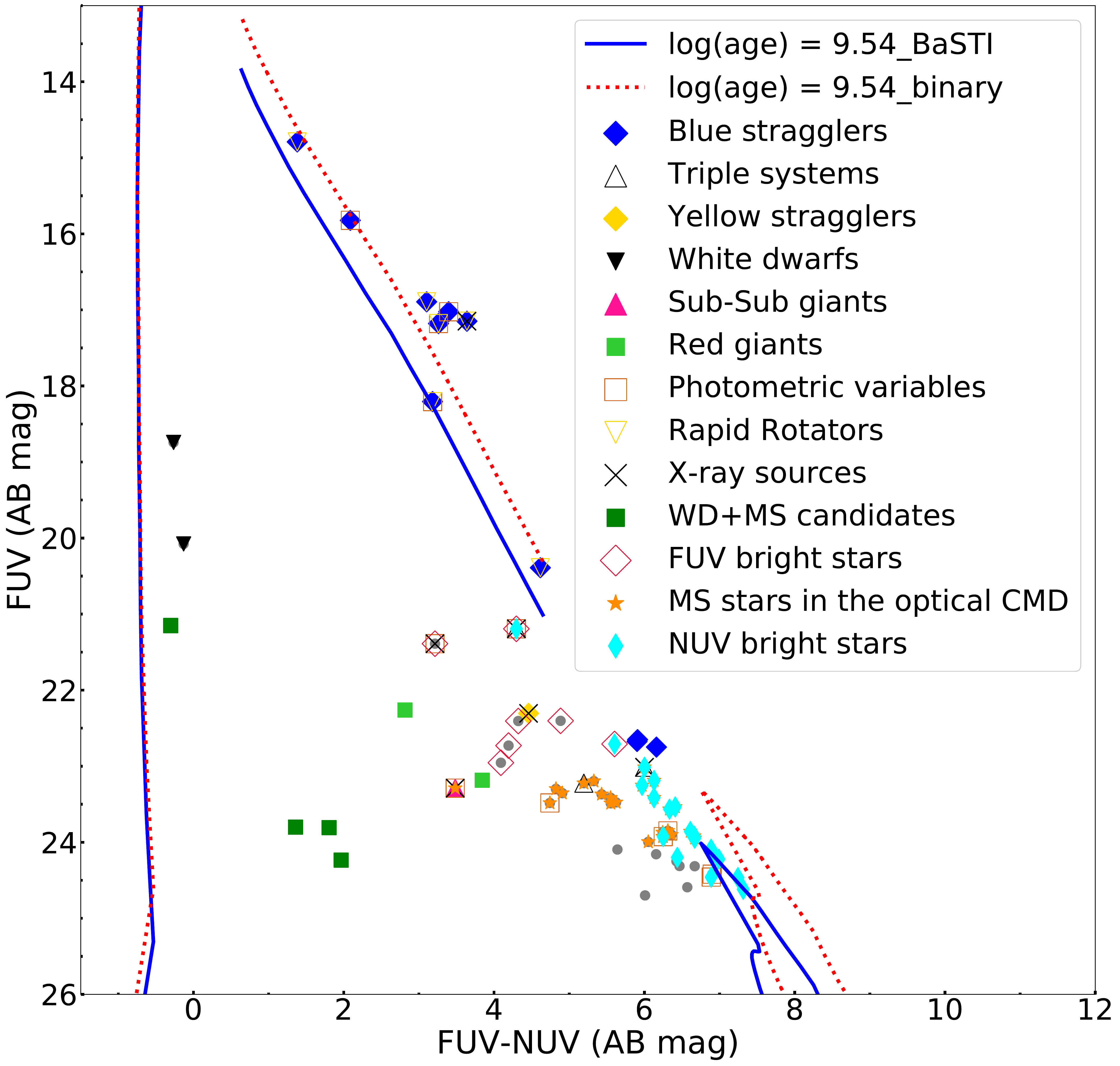}\includegraphics[width= 1.0\columnwidth]{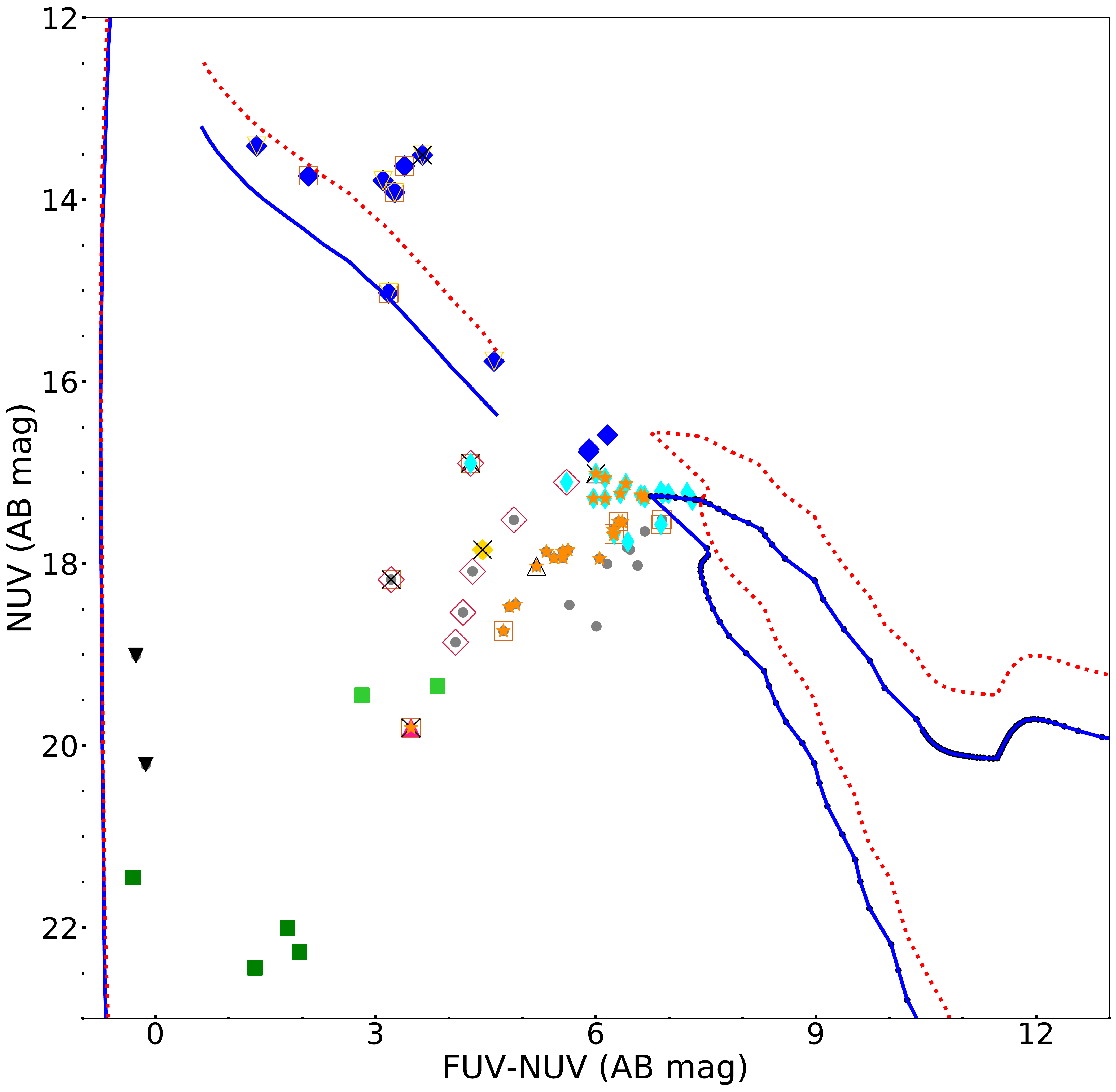}\\ 
  	\caption{(Top Left:) FUV, FUV$-$V CMD, (Top Right:) NUV, NUV$-$V CMD, (Bottom Left:) FUV, FUV$-$NUV CMD and (Bottom Right): NUV, FUV$-$NUV CMD of M67 is shown. A 3.5 Gyr UV isochrone is superposed on all CMDs along with the equal mass binary sequence. The WD cooling sequence is shown to the left of the MS and a BSS model line above the MSTO. The symbols have the same meaning in all the figures. Three regions (WD, Gap and BSS) are identified in the FUV, FUV$-$V CMD}
	\label{FUV-optical CMD}
\end{figure*}
An (FUV$-$V) vs. FUV CMD (henceforth 'FUV$-$V CMD') of M67 is shown in top left panel of Figure \ref{FUV-optical CMD}, and the (NUV$-$V) vs. NUV CMD (henceforth 'NUV$-$V CMD') is shown in top right panel of Figure \ref{FUV-optical CMD}. The bottom panels show the UV CMDs. An isochrone of 3.5 Gyr is superposed on all the CMDs along with the binary sequence. 

The FUV$-$V CMD guided by the overlaid isochrone indicates that only stars on and above the tip of the MSTO are detected in the FUV, as the tip of the MS is found at an FUV magnitude of $\sim$ 24. All these are found to be located above and on the MSTO, and upper MS region in the optical CMD. We observe that, 47 stars (orange filled stars \& pink open diamonds) appear brighter than the tip of MSTO in this CMD. Among these stars, 3 triple system, 8 X-ray sources, 1 SSG, and 9 photometric variables (PV) are identified from the literature. From the radial velocity study (\citealp{Geller2015}) it is found that 19 of them are binary stars and 28 are single member stars. The binary fraction found in this region is not inconsistent with the cluster binary fraction. Thus a variety of stars are found in this CMD and many have excess flux with respect to their location in the optical CMD. Note that some stars shown in the FUV$-$V CMD may not appear in the UV CMDs (shown in the lower panels), if they are not detected in the NUV, and vice versa. 
In the FUV$-$V CMD, we identify three regions corresponding to WDs, BSS and gap stars, as shown in the figure. The location of WDs and BSS are similar to those adapted in the Optical CMDs, whereas the gap region is similar to that found by \citet{Knigge2002}, in their FUV$-$optical CMD. The detected 3 WDs are located in the WD region. We identify 6 stars (filled green squares) in the gap region, which are found to be located in the MS, in the optical CMD. They are found on the bluer side of the FUV$-$V MS, suggesting that they may be hotter than expected from the optical CMD. The BSS region consists of stars brighter than the MS, along the BSS model line. We see that the BSS fall in this region. The two YSS also fall in this region. We notice a number of stars are found along with the bright BSS, where the model BSS line is also located. The region brighter than the tip of the MS and the fainter end of the BSS model line has three BSS along with a number of stars. This is an interesting point and we look into this in more detail below.

 In the  NUV$-$V vs. V CMD, shown in Figure \ref{FUV-optical CMD}, we notice a very well defined MS well fitted by an isochrone of 3.5 Gyr as well as the adopted reddening. Most of the stars brighter than 16.8 mag in the V band are detected in the NUV band. The stellar evolutionary features of the MS, MSTO, subgiant branch and the RGB stages are clearly delineated in the NUV$-$V CMD. The MS is found to be a tight sequence, with very little scatter in the (NUV$-$V) colour. Stars near the MSTO show a spread in magnitude in this CMD when compared to the optical CMD. Similar to the FUV$-$V CMD, BSSs appear as the brightest sequence in NUV$-$V CMD. 

The subgiant branch is clearly visible in the NUV$-$V CMD and it is well fitted by the isochrone (3.5 Gyr age). The RGB stars are detected in the NUV band and show a relatively large scatter. The isochrone more or less fits the bluer RGB stars, while the redder RGB stars tend to be brighter in the NUV and follow the binary isochrone. It is not clear whether the NUV excess detected among some RGB stars is due to binarity. Near the RGB region, 5 brightest stars form a clump in Figure \ref{FUV-optical CMD}. Among these stars, 4 of them are identified as RG clump stars by \citet{Sandquist2004}. Stars which appear in the FUV$-$V, NUV$-$V and the UV CMDs are discussed below. We first discuss the stars that are known from the optical CMD and then stars that are identified from the UV CMDs
\begin{table*}
\centering
\caption{List of M67 members detected by GALEX. Column 1 gives the ID from MMJ93, Column 2 gives the WOCS ID of M67. Column 3 \& 4 are V and B-V mag from MMJ93, Column 5 \& 6 are GALEX FUV \& NUV in AB mag, corrected for saturation as mentioned in the text. Column 7 are magnitude from UIT magnitude (Landsman et al. 1998). Chandra, ROSAT and XMM -Newton are the X-ray counterparts given in column 8, 9 \& 10. The membership and type/class taken from the radial velocity membership study are shown in column 11 \& 12. In the comments section, details regarding their UV properties are listed. (The full table is available online).}
\label{tab 1:Stars}
\begin{tabular}{lrcccccccccll} 
	\hline
	MMJ & WOCS & V & B-V & FUV & NUV & m$_{152}$ & CX & X &NX& Memb. &  Type/Class & Comments \\	
	\hline 
	\multicolumn{13}{c}{Blue stragglers}\\
	\hline 
	6490 & 1006 & 10.99 & 0.11 & 14.79 & 13.41 & 12.10 & ... & ... & ... & (BL)M & BSS, SB1, RR & \\
	6504 & 1007 & 10.94 & 0.22 & 17.18 & 13.92 & 14.47 & ... & ... & ... & BM & BSS, SB1, RR,EX Cnc & \\
	6511 & 1017 & 10.6 & 0.34 & 17.15 & 13.51 & 14.32 & 109 & ... & ... & (S)N & BSS, RR & \\
	5191 & 1020 & 12.7 & 0.48 & 22.67 & 16.77 & ... & ... & ... & ... & SM & BSS & \\
	6006 & 1025 & 12.28 & 0.39 & 20.39 & 15.77 & ... & ... & ... & ... & BM & BSS,SB1, RR & \\
	6510 & 1026 & 10.7 & 0.11 & 14.73 & ... & 12.06 & ... & ... & ... & (BL)M & BSS, SB1, RR & \\
	5699 & 2007 & 12.26 & 0.57 & 20.16 & ... & ... & ... & ... & ... & SM & BSS & \\
	6479 & 2011 & 11.28 & 0.13 & 15.82 & 13.74 & 13.15 & ... & ... & ... & SM & BSS, PV & \\
	... & 2013 & 10.92 & 0.31 & 16.89 & 13.79 & 14.20 & ... & ... & ... & BM & BSS, SB1, RR & \\
	6477 & 2015 & 12.04 & 0.60 & 22.65 & 16.74 & ... & ... & ... & ... & SM & BSS & \\
	\hline
	\multicolumn{13}{l}{SM - Single Member, SN-Single Non-member, BM- Binary member, BLM- Binary Likely Member,  U-unknown with RV measurements.}
\end{tabular}
\end{table*}
\subsubsection{Blue Straggler stars}
\label{sec:BSS}
 In the FUV band, 15 members and  WOCS 1017 BSSs are detected (listed in Table \ref{tab 1:Stars}). They appear as a bright sequence in the CMDs. The brightest and hottest BSS (WOCS 1010) in the optical CMD is saturated in both the FUV and NUV bands of GALEX, and hence could not be considered further in this study. We observe that BSSs span a range of $\sim$ 3 mag scale on the optical CMD; whereas, in the FUV$-$V CMD, they stretch over a range of  $\sim$ 8 mag region above the MSTO, and are easily identified. Among the 16 detected BSSs, 9 are relatively bright (brighter than 19 mag in FUV) and 7 of them are brighter than 17.2 mag. WOCS 1006 and WOCS 1026, which are single lined spectroscopic binaries (SB1) and rapid rotators (RR) are found to be brightest in the FUV ($\sim$ 14.7 mag). They are also among the brightest BSSs observed in the UIT images. In general, the BSSs that are brighter in the FUV are either RR and/or PV. Three BSSs (WOCS 1020, WOCS 2015 \& WOCS 2068) appear fainter than 22.5 mag in the FUV. Among the 16 BSSs detected in the FUV band, 4 (WOCS 1017, WOCS 2009, WOCS 4003 \& WOCS 5005) also have X-ray detection. WOCS 2009 is a triple system but also listed as a PV \& a double lined spectroscopic binary (SB2). WOCS 4003 and WOCS 5005 are relatively faint and are found closer to the base of the BSS model line in the FUV$-$V CMD (Figure \ref{FUV-optical CMD}). We find that 13 BSSs have NUV magnitudes, 8 are bright in the NUV, 5 are relatively fainter and they are located closer to the MSTO. There are 5 stars which are detected in the FUV but do not have the NUV magnitude. On the other hand, 2 stars (WOCS 3009, WOCS 9005) have the NUV detection but are not detected in the FUV band. WOCS 1026 is the brightest BSS observed in the FUV band, but its NUV magnitude could not be determined likely to be due to saturation in the GALEX observation. WOCS 1006 is the brightest in the NUV band. Overall BSSs dominate the UV light in old open clusters.
\subsubsection{Yellow Straggler stars }
\label{sec:YSS}
Three of the 4 known YSSs are detected in the FUV (WOCS 2002, WOCS 1015 and WOCS 2008), they are also SB1 stars. WOCS 2002 is relatively bright in the FUV, whereas the other two are relatively faint and have redder (FUV$-$V) colours. 2 YSSs are detected in the NUV band and are found to be located brighter than the subgiant branch in the NUV$-$V CMD. The brightest YSS, detected in the FUV which was identified in the BSS location, on the other hand, does not have NUV detection. All 3 stars have X-ray emission (ROSAT, Chandra and XMM-Newton). Two of these YSSs (WOCS 2002, WOCS 1015) are believed to have undergone mass transfer (\citealp{Landsman1997_S1040}, \citealp{Leiner2016}). WOCS 2002  is found to have a hot WD companion with a low mass of 0.23 M$\sun$ (\citealp{Landsman1997_S1040}). A recent study by \citet{Leiner2016} showed that WOCS 1015 has a normal MS or a BSS companion. They suggested that it is likely to be an evolved BSS.

 \subsubsection{Red giants}
 \label{sec:RG}
Two RGs (WOCS 1036, WOCS 1075) are detected in the FUV which are bright with redder colour in the optical CMD (Table \ref{tab 1:Stars}). The detection of these RGs, particularly in the FUV band, is quite intriguing since their (FUV$-$NUV) colour is $\lessapprox$ 4 mag and is bluer in the UV CMDs as seen in Figure \ref{FUV-optical CMD} (bottom panel). 
The FUV brightness as well as the blue UV colour are suggestive of extra flux in the UV when compared to the prediction based on the optical CMD.
 \subsubsection{Sub-Sub giants}
 \label{sec:SSG}
Some of the previous studies has often termed SSG as red stragglers. Recently \citet{Geller2017} clearly delineated them by giving their demographics  in the optical CMD, so as to separate SSG and red stragglers. Two SSGs are known in M67, both have high proper motion and radial velocity membership probability. Among two SSGs, only WOCS 15028 is detected in the FUV, which also has X-ray detection. The fact that this star, which is quite sub-luminous in the optical CMD, is detected in the FUV warrants attention. Both the stars are detected in the NUV band. They are fainter than the subgiant branch but bluer than the RGB stars. One SSG is relatively brighter and bluer in the NUV$-$V CMD, whereas both occupy the same location in the optical CMD.
\subsubsection{Triple system}
\label{sec:Triple systems}
 Four triple systems are detected in the FUV band (Table \ref{tab 1:Stars}) and 3 in the NUV band. WOCS 2009 is a BSS \& SB2 as discussed earlier and has the brightest FUV mag among the triple systems, and is not detected in the NUV band. WOCS 3012, WOCS 7008 and WOCS 4030 are the other three triple systems with FUV detection.WOCS 7008 and WOCS 4030 are identified at the FUV detection limit of GALEX. 
\subsubsection{White dwarfs}
\label{sec:WD}
There are 3 WD candidates (refer Table \ref{tab 1:Stars}) detected in the GALEX FUV, and 2 in the  NUV which were earlier identified in UIT data by \citet{landsman98}. They occupy the WD region in both the FUV$-$V and NUV$-$V CMDs. \citet{Fleming1997} derived the effective temperature of the 2 WDs; MMJ 5670 is the brightest WD candidate and is known as a hot DA WD with T$_{eff}$=68,230 $\pm$ 3200 K and MMJ 5973 to be a DB WD with T$_{eff}$=17,150 $\pm$ 150 K. MMJ 6061 is relatively fainter in the FUV.

\subsubsection{FUV bright stars}
\label{sec:FUV bright}
We identify 15 stars  located  in the same BSS region outlined in Figure \ref{FUV-optical CMD} along with the 16 BSSs, which are located on or below the MSTO, including the subgiant branch in the optical CMD. Thus, we classify these 15 stars as FUV bright stars (pink open diamonds) and are listed in Table \ref{tab 1:Stars}. These stars have FUV magnitude in the range 19.5 - 23.0, whereas the MSTO is at $\sim$ 24 mag. Among these 15,  6 are significantly FUV bright (WOCS 2012, WOCS 11005, WOCS 3012, WOCS 6008, WOCS 3001 and WOCS 11006) based on the FUV$-$V CMD (Figure \ref{FUV-optical CMD}). These 6 stars are located close to the BSS model line as shown in the FUV$-$V CMD, with FUV $<$22 mag, at least 2 magnitude  brighter than the MSTO.  Among the 6 stars, 5 stars have high membership probability based on both proper motion \& radial velocity measurements. WOCS 11006 is a member based on proper motion study. WOCS 2012 is the brightest among these 6 stars and lie close to the BSS model line (Figure \ref{FUV-optical CMD}), though it lies on the subgiant branch in the optical CMD. WOCS 6008 and WOCS 11005 are subgiant branch stars in the optical CMD. Among these 15, 9 stars are in the FUV magnitude range of 22.0 - 23.0, fainter than the above 6, but brighter than the MSTO. These are identified close to the 3 FUV faint BSSs ($\sim$ 22.7 mag). Among these 9 stars, \citet{Sandquist2004} suggested that WOCS 8004 and WOCS 6006 are likely BSS candidates. In the UV CMDs, the FUV bright stars shift to a bluer region in the CMD, when compared to their location in the FUV$-$V CMD, suggesting they are brighter and probably hotter in the FUV. It can also be seen that in the NUV$-$FUV vs NUV CMD, they occupy a region fainter than the MSTO, suggesting that they are not bright in the NUV. A detailed spectral energy analysis using multi-wavelength observation data of all these FUV bright stars are presented in section \ref{SED}.

\subsubsection{White dwarf + Main sequence binaries}
\label{sec:WD+MS}
Apart from the 3 WD candidates, in the FUV$-$V CMD (Figure \ref{FUV-optical CMD}), we identify 6 possible WD+MS binaries (6 green filled squares in the gap region of Figure \ref{FUV-optical CMD}). These stars are identified based on the fact that they occupy MS location in the optical CMD, well below the MSTO (more than 2 magnitude fainter than MSTO), whereas they are brighter and bluer than the MSTO stars in the UV CMD, in a region between the MS and WD cooling sequence. These are similar to WD+MS binaries shown by \citet{Parsons2016} in their figure 1 and 2. The UV excess could be possibly due to a WD companion to the low mass GK type MS stars, similar to the population of FGK type stars with WD companion identified by \citet{Parsons2016}. \citet{Knigge2002}, while studying the FUV observations of 47 Tuc, suggested a cataclysmic variable (CV) zone in their FUV-optical CMD. The other stellar types which could occupy this region, other than CVs, are MS+WD binaries and He WDs. This region is similar to the location of gap stars, found by \citet{Haurberg2010}, \citet{Dieball2010_M80}  and \citet{Dieball2017_NGC6397} in their UV study of M15, M80 and NGC 6397 respectively. They suggested the region between WD and MS in the FUV$-$NUV CMD belong to gap stars, which could be a combination of CVs, detached WD+MS binaries and He WDs. \citet{Hurley2005} based on the N-body model of M67, expected to find 226 single WDs, 60 double WD binaries and 145 WD in non-WD companion near the cluster centre. They also expected to find another 33 WDs in other binaries, some with low mass MS, which could appear near the WD sequence and suggest that the observation of WDs in M67 is incomplete.
Similar to the 6 stars which lie to the bluer side of FUV$-$V CMD,  we observe another 5 stars which are significantly ($\sim$ 1 magnitude) bluer than the MS stars but are redder than the WD in the NUV$-$V CMD, they are listed in Table \ref{tab 1:Stars}. Further analysis of these objects are presented in section \ref{SED}.
 \subsubsection{Anomalous stars in the NUV$-$V CMD}
 \label{sec:Anomalous NUV stars}
  An inspection of the NUV$-$V CMD reveals a group of stars (cyan filled diamonds in Figure \ref{FUV-optical CMD}) in the NUV magnitude range of 16.5 to 17.3. This group consisting of 31 stars are located between the MSTO of the single and binary star isochrone, and we refer to them as NUV bright stars. These stars occupy the MSTO in the optical CMD in the Figure \ref{FUV-optical CMD}, thus they show a moderate excess in the NUV flux (the excess is up to 0.5 mag, whereas the photometric error is less than 0.1 mag). Among these, 2 stars belong to the FUV bright stars which show large excess (WOCS 3012, WOCS 3015) and a few stars have less FUV excess (orange filled star) in Figure \ref{FUV-optical CMD} (upper left panel). From the radial velocity estimation, 23 stars are classified to be single members and 8 stars are binary members. Of the 8 binary stars; 3 (WOCS 4004, WOCS 3023, WOCS 4051) are known SB1 while 5 (WOCS 3012, WOCS 7008, WOCS 6010, WOCS 3015, WOCS 3006) are known SB2 stars. In short, these NUV bright stars appear to be populated by various types of stars. Some show only NUV excess, whereas some show FUV excess as well. Five of the NUV bright stars are classified as MSTO gap stars by \citet{Sandquist2004}. They suggest that these stars are likely to be categorised as BSSs. 

In Figure \ref{FUV-optical CMD}, we identify another 9 stars (cyan filled diamonds) which are found on the bluer side of the MSTO. They are also NUV bright stars, but in the optical CMD, they are located below the MSTO. All of them have radial velocity membership probability $>$ 90\%, but only 7 stars have proper motion membership probability $>$ 99\%. Among these stars, 5 are single and 4 are binary members. WOCS 5035 is a PV and classified as a MSTO gap star by \citet{Sandquist2004}. WOCS 8007, WOCS 11022 and WOCS 2016 are SB1, while WOCS 4016 is an SB2 star. \citet{Sandquist2004} suggested that WOCS 7044 and WOCS 7015 could be tentatively classified as BSSs based on their VI CMD. In the optical CMD, these stars are found at least 1 magnitude below the MSTO. 
 
  In the NUV$-$V CMD, we identify a star (WOCS 2083) which has a radial velocity membership of 97\%, and is located below the RGs and redder than the MS in Figure \ref{FUV-optical CMD}. This star is not shown in the optical CMD since the MMJ93 does not provide B magnitude. Taking the value of B = 13.52 from \citet{Sanders1977} and V =12.49 from MMJ93, the star will lie on the subgiant region in the optical CMD. We suggest that it could be an SSG candidate. 
\subsection{X-ray counterparts}
\label{X-ray counterparts}
There are 36 optical member stars with X-ray observation in at least one of the three missions mentioned earlier. We have cross identified GALEX UV sources with their X-ray counterparts. In Table \ref{tab 1:Stars} we have listed 25 stars which have both UV and X-ray detection. Among them, 15 sources are detected in the FUV, 16 in the NUV and 6 in both the filters. Among these stars, 20 are binary members, 2 are single stars, 3 are unknown and WOCS 1017 of the cluster based on the radial velocity study.

 Four BSSs are detected in both X-ray and GALEX, as discussed in section \ref{sec:BSS}. WOCS 4011 is an RGB star with an X-ray luminosity of the order of 10$^{30}$ ergs/s and is detected only in the NUV. This star is a single member of the cluster from the radial velocity study. It is recently found to host a Jupiter-mass exoplanet (\citealp{Brucalassi2014}). Two SSGs were detected in ROSAT (\citealp{Belloni98}), and both have X-ray luminosity of 7.3$\times$10$^{30}$ ergs/s. \citet{Mooley2015} studied the X-ray spectrum and variability of one SSG (WOCS 13008) and suggested that the X-ray emission is from its corona. \citet{Leiner2017} studied both these stars and stated that the X-rays are from the hot corona, due to strong magnetic field. Recently \citet{Geller2017_SSG} suggested a few formation pathways for these stars.  

 Three stars (WOCS 2009, WOCS 3012 and WOCS 7008) are in triple systems and are found to be RS CVn systems. All three stars are bright in the FUV and are discussed in section \ref{sec:Triple systems}. WOCS 6010 lies on the MS in the optical CMD and is also located in the NUV bright star region in the NUV$-$V CMD. The star is detected only in Chandra and has an X-ray luminosity of 4.2$\times$10$^{28}$ ergs/sec, though \citet{VandenBerg2004_Chandra} stated that the X-ray and optical position are uncertain due to relatively large offset. WOCS 10025 could be a possible CV based on the X-ray spectrum suggested by \citet{Belloni98}, though \citet{Mooley2015} contradict this possibility. This star is detected only in the NUV and is located on the MS, close to the binary sequence in the NUV$-$V CMD. 
 
  Three YSSs have X-ray detections in all the three missions. WOCS 2008 and WOCS 1015 are redder than the MS in the FUV$-$V CMD. \citet{Mooley2015} indicated that tidal interaction cannot explain the X-ray emission of both these stars due to their wide separation. In the NUV$-$V CMD, only 1 (WOCS 1015) YSS with X-ray counterpart is detected. Four contact binaries, which are all W Uma type (\citealp{Mooley2015}) are detected in the UV. Among these 4 contact binaries, 2 are detected only in the FUV and 2 only in the NUV.

 \subsection{Spectral Energy Distribution (SED) of UV bright stars}
\label{SED}
The BSSs are found to have a large range in FUV flux, in comparison to the optical flux. It is therefore essential to identify the parameters of BSSs which makes them more sensitive in the FUV. Also the FUV bright stars are interestingly found to be co-located with the known BSSs and will be good to compare their properties. The fundamental properties, which can be estimated and compared are the total Luminosity (L/L$_{\sun}$), the radius (R/R$_{\sun}$) and the effective temperature (T$_{eff}$). We derive these parameters for the BSSs as well as the {FUV bright stars} using Spectral Energy Distribution (SED). SEDs are constructed with the observed photometric data from UV-to-IR and fitted with Kurucz models (\citealp{Castelli1997}). SED is built using a virtual observatory-tool, Virtual Observatory SED analyser (VOSA) (\citealp{Bayo2008}). We have used the existing photometric data from UV (GALEX - [FUV, NUV]), optical (MMJ93 - [U, B, V, R, I], GAIA - [G]), near-infrared (2MASS - [J, H, K$_{s}$]) to mid-infrared (WISE - [W$_{1}$, W$_{2}$, W$_{3}$]). GAIA, 2MASS and WISE data are obtained from VO photometry (table \ref{tab:Photometry}). The crossmatch is found within a spatial search radius of 3\arcsec. The input file to the tool consists of stellar parameters such as RA, Dec, distance, flux, filter etc. The tool corrects the observed flux for extinction (\citealp{Fitzpatrick1999}, \citealp{Indebetouw2005}) in the respective wavelength band. It utilises the filter transmission curve to calculate the synthetic photometry of Kurucz model. This tool then allows the user to construct an SED and perform statistical tests to compare the observed data with the synthetic photometry computed from the model. The tool estimates T$_{eff}$, surface gravity (g), luminosity (L/L$_{\sun}$) and stellar radius (R/R$_{\sun}$) along with the errors for a given fit. Details of the fit are provided in \citet{2016Subramaniam}. Radius is estimated by using the scaling factor M$_{d}$, which is given by ${M}_{d}={\left(\frac{R}{D}\right)}^{2}$, where R is the radius and D is the distance to the cluster (same as that mentioned in section 2.3). Stellar mass is determined using the Seiss theoretical isochrones and evolutionary tracks (\citealp{Siess2000}), after obtaining the T$_{eff}$ and L/L$_{\sun}$ values from the fit. The evolutionary tracks underestimate the mass of the BSSs by 15\%, as seen in two BSSs in NGC 188 (\citealp{Mathieu2009Natur}, \citealp{Geller2011Natur}). We estimate the mass only to compare the mass range for BSSs and FUV bright stars.

\begin{figure*}
	\includegraphics[width=2.0\columnwidth]{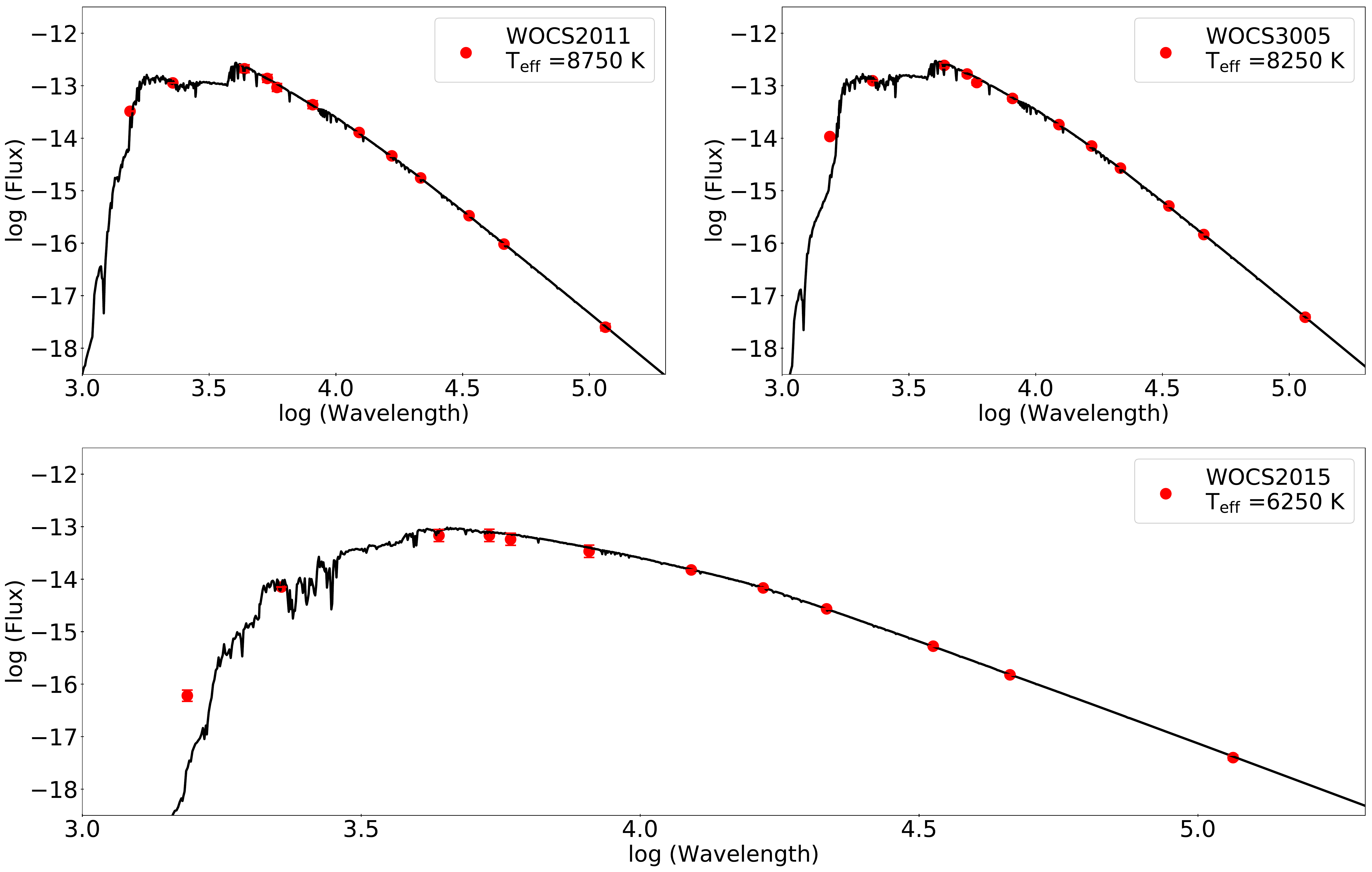}

	\caption{The SEDs (extinction corrected) of WOCS 2011, WOCS 3005 and WOCS 2015 with photometric flux from UV to IR. Best fitting Kurucz model spectrum is overplotted and the corresponding temperature is listed in each panel. The unit of wavelength is \angstrom\, and flux is \flux.}
	\label{SED of BSS}
\end{figure*}

\begin{table*}
	\caption{The photometry of stars, for which we have constructed SEDs. U, R and I are taken from MMJ93. 2MASS, WISE and GAIA are taken from their respective point source catalogue through VO photometry. Magnitudes in the other wavelength are listed in table \ref{tab 1:Stars}.}
	\label{tab:Photometry}
	\begin{tabular}{lllllllllll}
		
		\hline
		Star Name & U & R & I & WISE-1 & WISE-2 & WISE-3 & 2MASS-J & 2MASS-H & 2MASS-Ks & Gaia-G\\
		\hline
		WOCS2011 & --- & --- & 11.1 & 10.97$\pm$0.02 & 10.99$\pm$0.02 & 11.07$\pm$0.14 & 11.02$\pm$0.02 & 11.01$\pm$0.02 & 10.99$\pm$0.02 & 11.24\\
		WOCS3005 & --- & --- & 10.8 & 10.50$\pm$0.03 & 10.54$\pm$0.02 & 10.57$\pm$0.09 & 10.65$\pm$0.02 & 10.54$\pm$0.02 & 10.53$\pm$0.02 & 11.01\\
		WOCS2015 & --- & --- & 11.37 & 10.47$\pm$0.02 & 10.52$\pm$0.02 & 10.57$\pm$0.09 & 10.85$\pm$0.02 & 10.59$\pm$0.02 & 10.52$\pm$0.02 & 11.76\\
		WOCS2003 & 13.20$\pm$0.01 & 12.18 & 11.84 & 10.94$\pm$0.02 & 11.02$\pm$0.02 & 11.09$\pm$0.13 & 11.43$\pm$0.02 & 11.15$\pm$0.02 & 11.12$\pm$0.02 & 12.32\\
		WOCS2012 & 13.45$\pm$0.01 & --- & 11.95 & 11.09$\pm$0.02 & 11.10$\pm$0.02 & 10.49$\pm$0.08 & 11.49$\pm$0.02 & 11.22$\pm$0.02 & 11.19$\pm$0.02 & 12.44\\
		WOCS3001 & 13.7$\pm$0.02 & 12.98 & 12.67 & 12.06$\pm$0.02 & 12.03$\pm$0.03 & 12.04$\pm$0.32 & 12.36$\pm$0.02 & 12.16$\pm$0.02 & 12.08$\pm$0.02 & 13.10\\
		WOCS3015 & 13.3 & --- & 12.02 & 11.24$\pm$0.02 & 11.27$\pm$0.02 & 11.39$\pm$0.17 & 11.60$\pm$0.02 & 11.36$\pm$0.02 & 11.28$\pm$0.02 & 12.50\\
		WOCS3024 & 14.30$\pm$0.02 & --- & 13.03 & 12.07$\pm$0.02 & 12.11$\pm$0.02 & 11.99$\pm$0.3 & 12.58$\pm$0.02 & 12.34$\pm$0.03 & 12.26$\pm$0.03 & 13.46\\
		WOCS5030 & 14.01 & --- & --- & 12.04$\pm$0.02 & 12.09$\pm$0.03 & 12.04$\pm$0.33 & 12.40$\pm$0.02 & 12.14$\pm$0.03 & 12.10$\pm$0.02 & 13.29\\
		WOCS6006 & 13.42 & --- & 12.24 & 11.49$\pm$0.02 & 11.52$\pm$0.02 & 11.71$\pm$0.23 & 11.80$\pm$0.02 & 11.57$\pm$0.02 & 11.52$\pm$0.02 & 12.65\\
		WOCS6008 & 13.54 & 12.33 & 11.92 & 10.92$\pm$0.02 & 10.98$\pm$0.02 & 11.06$\pm$0.13 & 11.38$\pm$0.02 & 11.08$\pm$0.02 & 11$\pm$0.02 & 12.43\\
		WOCS7009 & 13.97$\pm$0.03 & --- & 12.63 & 12.03$\pm$0.02 & 12.05$\pm$0.03 & 12.17$\pm$0.35 & 12.39$\pm$0.02 & 12.15$\pm$0.02 & 12.09$\pm$0.02 & 13.18\\
		WOCS7010 & --- & --- & 12.07 & 11.27$\pm$0.02 & 11.30$\pm$0.02 & 11.53$\pm$0.20 & 11.67$\pm$0.02 & 11.38$\pm$0.02 & 11.34$\pm$0.02 & 12.54\\
		WOCS8004 & 13.81$\pm$0.02 & 12.85$\pm$0.02 & 12.52 & 11.77$\pm$0.02 & 11.79$\pm$0.02 & 11.87$\pm$0.27 & 12.13$\pm$0.02 & 11.91$\pm$0.02 & 11.85$\pm$0.02 & 12.98\\
		WOCS11005 & 13.34$\pm$0.01 & --- & 12.07 & 10.76$\pm$0.02 & 10.85$\pm$0.02 & 10.75$\pm$0.10 & 11.28$\pm$0.02 & 10.90$\pm$0.03 & 10.82$\pm$0.02 & 12.47\\
		WOCS11006 & 13.74 & --- & 12.69 & 10.94$\pm$0.02 & 11.00$\pm$0.02 & 10.88$\pm$0.11 & 11.50$\pm$0.02 & 11.09$\pm$0.02 & 11.02$\pm$0.02 & 12.71\\
		WOCS17028 & 14.75 & --- & 13.4 & 12.51$\pm$0.02 & 12.58$\pm$0.03 & 11.95$\pm$0.26 & 13.02$\pm$0.02 & 12.72$\pm$0.03 & 12.67$\pm$0.02 & 13.93\\
		WOCS1036 & --- & --- & --- & 4.99$\pm$0.08 & 4.93$\pm$0.03 & 4.99$\pm$0.02 & 6.01$\pm$0.02 & 5.26$\pm$0.02 & 5.02$\pm$0.02 & 8.10\\
		WOCS1075 & --- & --- & 6.68 & 4.22$\pm$0.10 & 3.84$\pm$0.06 & 4.26$\pm$0.01 & 5.39$\pm$ 0.05 & 4.69$\pm$0.04 & 4.36$\pm$0.04 & 7.81\\
		WOCS17029 & --- & --- & --- & 13.47$\pm$0.03 & 13.56$\pm$0.04 & --- & 14.09$\pm$0.03 & 13.63$\pm$0.02 & 13.51$\pm$0.03 & 15.38\\
		WOCS19032 & 17.77 & --- & 14.9 & 13.27$\pm$0.03 & 13.37$\pm$0.04 & --- & 14.07$\pm$0.03 & 13.58$\pm$0.04 & 13.43$\pm$0.04 & 15.60\\
		WOCS19045 & --- & --- & --- & 14.33$\pm$0.03 & 14.36$\pm$0.07 & --- & 14.76$\pm$0.03 & 14.50$\pm$0.05 & 14.36$\pm$0.05 & 15.76\\
		WOCS21027 & 16.95 & --- & 15.66 & 14.88$\pm$0.04 & 14.86$\pm$0.09 & --- & 15.21$\pm$0.04 & 14.97$\pm$0.07 & 14.76$\pm$0.08 & 16.16\\
		WOCS23028 & 16.8 & --- & 15.5 & 13.42$\pm$0.03 & 13.53$\pm$0.04 & --- & 13.98$\pm$0.03 & 13.59$\pm$0.03 & 13.51$\pm$0.03 & 15.16\\
		WOCS24022 & 16.31 & --- & 14.34 & 13.32$\pm$0.03 & 13.37$\pm$0.04 & --- & 13.81$\pm$0.02 & 13.44$\pm$0.03 & 13.34$\pm$0.03 & 14.93\\
		WOCS36035 & --- & --- & 15.46 & 14.8$\pm$0.04 & 14.8$\pm$0.10 & --- & 15.08$\pm$0.04 & 14.81$\pm$0.07 & 14.71$\pm$0.07 & 15.88\\
		MMJ6427 & 19.08 & --- & 15.78 & 14.03$\pm$0.03 & 14.15$\pm$0.05 & --- & 14.82$\pm$0.04 & 14.22$\pm$0.05 & 14.05$\pm$0.04 & 16.45\\
		MMJ5658 & 18.75$\pm$0.06 & 15.88 & 15.19 & 13.36$\pm$0.03 & 13.37$\pm$0.04 & --- & 14.33$\pm$0.03 & 13.79$\pm$0.03 & 13.56$\pm$0.04 & 16.01\\
		MMJ6398 & 17.35 & --- & 15.91 & 14.88$\pm$0.04 & 14.88$\pm$0.09 & --- & 15.46$\pm$0.05 & 15.13$\pm$0.07 & 15.11$\pm$0.12 & 16.46\\
		MMJ6409 & --- & --- & 16.69 & 14.57$\pm$0.04 & 14.74$\pm$0.08 & --- & 15.59$\pm$0.07 & 14.82$\pm$0.06 & 14.63$\pm$0.07 & ---\\
		\hline
	\end{tabular}
\end{table*}
We fitted SEDs for 18 BSSs (those detected in GALEX), of which we present the results of 17 BSSs. We have excluded the BSS (WOCS 2009), which is in a triple system. The estimated parameters T$_{eff}$, R/R$_{\sun}$, L/L$_{\sun}$, and M/M$_{\sun}$ along with the errors of 17 BSSs are tabulated in Table \ref{tab 5:BSS_SED}. These are the values corresponding to the best fitting Kurucz model spectrum. Though we estimated log(g) values, we do not tabulate our estimations, as SED is not the ideal method to estimate log(g). In general, the estimated values of log(g) range between 3-5 dex. The estimated temperatures are compared with previous measurements (SED fitting method) by \citet{Deng99}, \citet{Liu2008_BSS} as listed in the Table \ref{tab 5:BSS_SED}. We have shown the SED fit of three BSSs - WOCS 2015, WOCS 2011 and WOCS 3005 in Figure \ref{SED of BSS} to demonstrate our fitting. An inspection of the table reveals that the BSSs have a large range in both T$_{eff}$ ($\sim$ 6000 - 9000 K) and L/L$_{\sun}$ ($\sim$ 5-30), and a small range in R/R$_{\sun}$ ($\sim$ 1.5 - 3.0). The estimated temperatures compare well with those from the earlier studies. 

\citet{Gilliland1992} determined the parameters of two stars (WOCS 4006 \& WOCS 1007) as L/L$_{\sun}$= 26.7, M/M$_{\sun}$=2.13, R/R$_{\sun}$= 2.8 and T$_{eff}$=7900 K for WOCS 1007 and L/L$_{\sun}$= 8.32, M/M$_{\sun}$= 1.65,  R/R$_{\sun}$= 1.52 and T$_{eff}$=7706 K for WOCS 4006. \citet{Bruntt2007} analysed these two variable stars and estimated the mass and temperature using theoretical pulsation models. From their estimation the mass range for WOCS 4006 was found to be 1.7 - 2.0 M$_{\sun}$ for T$_{eff}$ 7930 - 8210 K and for WOCS 1007, it is found to be 1.8 - 2.3 M$_{\sun}$  for T$_{eff}$ 7450-7740 K. Our estimated values of mass and temperature for these stars using SED fitting are comparable with the estimations of \citet{Bruntt2007} and \citet{Gilliland1992}, though we derive slightly lower luminosities for these stars. 

\begin{table*}
	\centering
	\caption{The parameters of BSSs from the SEDs. Column 1 gives ID from MMJ93, Column 2 gives the WOCS ID of M67, Column 3 gives the temperature derived from SED fitting, Column 4 \& 5 give the temperature from literature (Deng et al 1999 and Liu et al 2008). Column 6, 7 \&  8 give the radius, mass and luminosity in solar units. }
	\label{tab 5:BSS_SED}
	\begin{tabular}{lclccccl} 
		\hline
			MMJ & WOCS & T$_{eff}$ (K) &  T$^{a}_{eff}$ (K)&  T$^{b}_{eff}$ (K)& Radius (R$\sun$)  & Mass(M$\sun$) & Luminosity (L$\sun$) \\
		\hline	
		6490 & 1006 & 9000 $\pm$ 125  & 8000 & 8950 & 2.11 $\pm$ 0.028   & 2.20	& 27.60 $\pm$ 0.74  \\
		6504 & 1007 & 7750 $\pm$ 125  & 6750 & 7900 & 2.88 $\pm$ 0.038   & 2.00 & 22.29 $\pm$ 0.79 \\
		6511 & 1017 & 8000 $\pm$ 125  &           &      & 2.93 $\pm$ 0.04   & 2.19 & 29.52 $\pm$ 0.78  \\
		5191 & 1020 & 6750 $\pm$ 125  & 6250 & 6500 & 1.75 $\pm$ 0.023   & 1.55 &  4.96 $\pm$ 0.14   \\
		6006 & 1025 & 7250 $\pm$ 125  & 6500 & 7050 & 1.84 $\pm$ 0.024   & 1.60 &  6.63 $\pm$ 0.18  \\ 
		6510 & 1026 & 8750 $\pm$ 125  & 8000 & 8500 & 2.54 $\pm$ 0.035   & 2.20 & 34.67 $\pm$ 1.44   \\
		5699 & 2007 & 6000 $\pm$ 125  & 6000 & 6200 & 2.66 $\pm$ 0.035   & 1.66 &  8.20 $\pm$ 0.23  \\
	    6479 & 2011 & 8750 $\pm$ 125  & 8000 & 8450 & 1.94 $\pm$ 0.026   & 2.14 & 19.72 $\pm$ 0.54  \\
	    	& 2013 & 8000 $\pm$ 125  & 7250 & 8100 & 2.78 $\pm$ 0.037   & 2.00 & 22.56 $\pm$ 0.71 \\
		6477 & 2015 & 6250 $\pm$ 125  & 6000 & 6150 & 2.88 $\pm$ 0.038   & 1.58 & 10.42 $\pm$ 0.29  \\
		& 2068 & 6500 $\pm$ 125  & 6000 & 6450 & 2.39 $\pm$ 0.032   & 1.60	&  8.05 $\pm$ 0.23   \\
		6501 & 3005 & 8250 $\pm$ 125  & 7000 & 8050 & 2.45 $\pm$ 0.032   & 2.01 & 20.89 $\pm$ 0.56 \\
		6047 & 3009 & 6250 $\pm$ 125  & 6000 & 6950 & 2.45 $\pm$ 0.032   & 1.62	&  8.08 $\pm$ 0.22\\     
		5833 & 4003 & 6750 $\pm$ 125  & 6250 & 6500 & 1.75 $\pm$ 0.023   & 1.52&  4.84 $\pm$ 0.16	\\
		5940 & 4006 & 8000 $\pm$ 125  & 6750 & 7800 & 1.50 $\pm$ 0.020    & 1.77&  6.74 $\pm$ 0.18\\
		5667 & 5005 & 6500 $\pm$ 125  & 6250 & 6600 & 2.37 $\pm$ 0.031   & 1.60 &  8.42 $\pm$ 0.30 \\ 
		5571 & 9005 & 6500 $\pm$ 125  & 6250 & 6100 & 1.87 $\pm$ 0.025   & 1.48&  5.37 $\pm$ 0.16\\
		\hline
		\multicolumn{8}{l}{$^a$ \citet{Deng99}, the temperature is taken from their Kurucz model fit.}\\
		\multicolumn{8}{l}{$^b$  \citet{Liu2008_BSS}. the temperature is taken from their Bluered model fit.}\\
\end{tabular}
\end{table*}

\begin{figure*}
	\includegraphics[width=2.0\columnwidth]{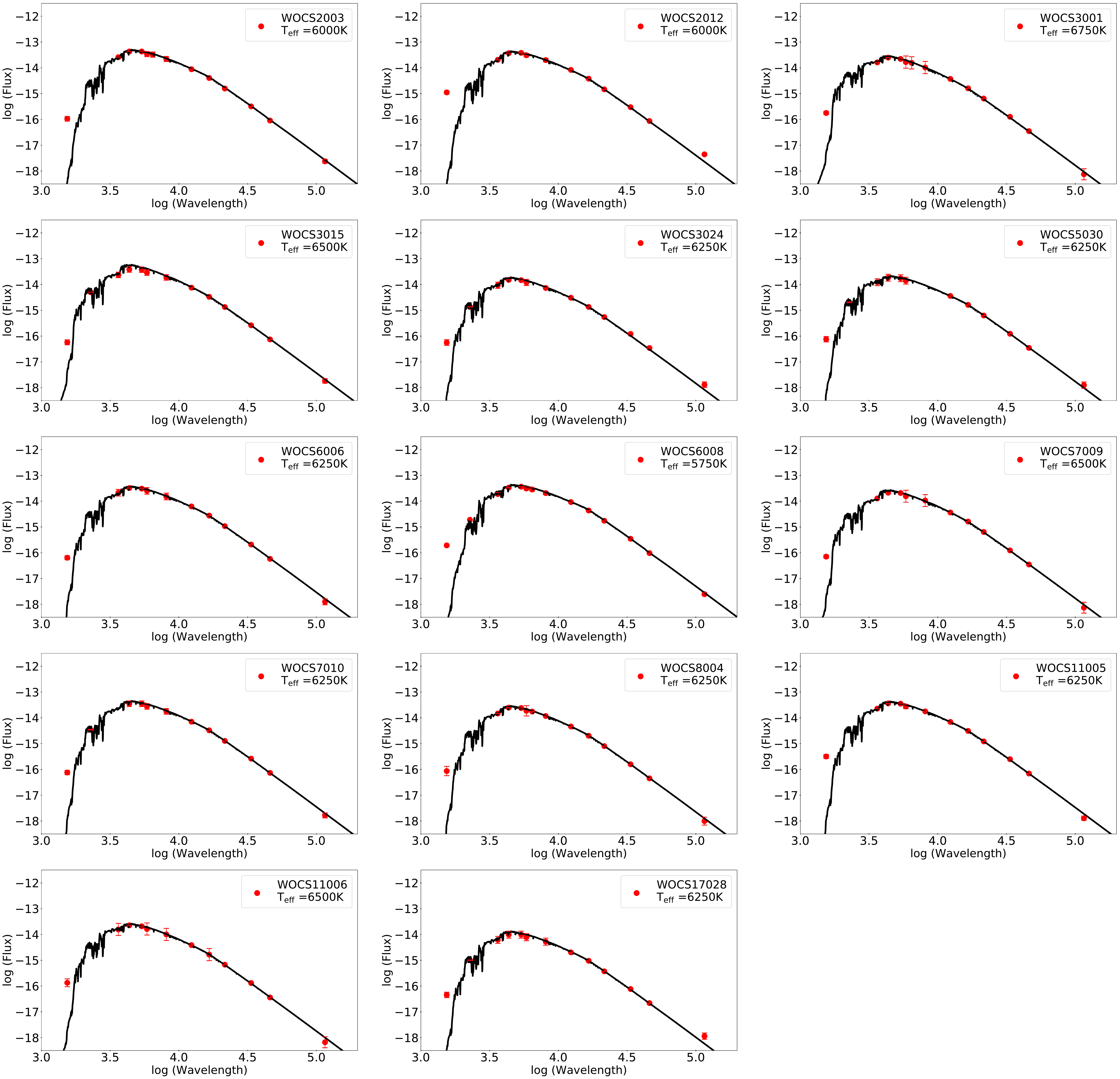}
		\caption{The SEDs (extinction corrected) of FUV bright stars with photometric flux from UV to IR. Best fitting Kurucz model spectrum is overplotted and the corresponding temperature is listed in each panel. The unit of wavelength is \angstrom\, and flux is \flux.}
	\label{SED of BSS candidates}
\end{figure*}

We also constructed SEDs for 15 FUV bright stars which are shown in Figure \ref{SED of BSS candidates}. The spectral fit to the SEDs reveal that we are able to fit with one spectrum for all the candidates, except WOCS 3012, which is known to be a triple system. From the figure, it is clear that all of these stars show certain amount of excess flux in the FUV, similar to that noticed in the SED of WOCS 2015. In the case of WOCS 2012 and WOCS 6008, we notice relatively large excess in the FUV flux, where the observed flux is about two orders more than the expected flux. The SEDs are used to estimate their T$_{eff}$, R/R$_{\sun}$, L/L$_{\sun}$, and M/M$_{\sun}$, which are listed along with the error in Table \ref{tab 6:BSS_candidates_SED}. \citet{Yakut2009} estimated the parameters (L/L$_{\sun}$= 2.78, R/R$_{\sun}$= 1.40, M/M$_{\sun}$= 1.47 and T$_{eff}$ = 6300 K) of WOCS 7009 in their table 13, which is an AH Cnc, a contact binary. \citet{Peng2016} also estimated radius (R/R$_{\sun}$= 1.33) and mass (M/M$_{\sun}$= 1.18) for WOCS 7009. Our estimation of L/L$_{\sun}$, R/R$_{\sun}$ and T$_{eff}$ are in good agreement with the estimates using other methods. When we compare the properties of the BSSs with these FUV bright stars, we find these span a smaller range in following three parameters; T$_{eff}$ ($\sim$ 5750 - 6750 K); R/R$_{\sun}$ ($\sim$ 1.0 - 2.5) and L/L$_{\sun}$ ($\sim$ 1.5 - 6.3). The FUV bright stars have the lower end of the parameter range of the BSSs.
	\begin{table*}
	\centering
	\caption{The parameters of FUV bright stars from the SEDs. Column 1 gives ID from MMJ93, Column 2 gives the WOCS ID of M67,  Column 3 gives the temperature derived from SED fitting, Column 4, 5 \&  6 give the radius, mass and luminosity in solar units.}
	\label{tab 6:BSS_candidates_SED}
	\begin{tabular}{lcllcl} 
		\hline
		MMJ & WOCS & T$_{eff}$ & Radius (R$\sun$)  & Mass(M$\sun$) & Luminosity (L$\sun$)	\\
		\hline
		5654 & 2003 & 6000 $\pm$ 125 & 2.34 $\pm$ 0.032 &	1.40&		6.32 $\pm$	0.24 \\
		5388 & 2012	& 6000 $\pm$ 125 & 2.18 $\pm$ 0.030 &	1.51 &		5.61 $\pm$	0.17 \\
		5741 & 3001	& 6750 $\pm$ 125 & 1.31 $\pm$ 0.018 &	       &		2.96 $\pm$	0.10 \\
		5451 & 3012	& 6500 $\pm$ 125 & 1.90 $\pm$ 0.026 &	1.50&		6.22 $\pm$	0.18\\
		6076 & 3015	& 6500 $\pm$ 125 & 2.00 $\pm$ 0.028 &	1.47 &		5.30 $\pm$	0.16\\
		5249 & 3024	& 6250 $\pm$ 125 & 1.28 $\pm$ 0.018 &	      &		2.16 $\pm$	0.06\\
		6467 & 5030	& 6250 $\pm$ 125 & 1.39 $\pm$ 0.019 &	1.30  &		2.63 $\pm$	0.09\\
		5969 & 6006	& 6250 $\pm$ 125 & 1.81 $\pm$ 0.025 &	1.40	&		4.46 $\pm$	0.15\\
		5993 & 6008	& 5750 $\pm$ 125 & 2.47	$\pm$ 0.034 &	1.69&		5.93 $\pm$	0.17\\
		6027 & 7009	& 6500 $\pm$ 125 & 1.35 $\pm$ 0.019 &	1.39&    	2.74 $\pm$	0.09\\ 
		5825 & 7010	& 6250 $\pm$ 125 & 1.99	$\pm$ 0.027 &	1.41 &		5.11 $\pm$	0.16\\
		5603 & 8004	& 6250 $\pm$ 125 & 1.58 $\pm$ 0.022 &	1.30    	&	    3.35 $\pm$	0.11\\
		5675 & 11005& 6250 $\pm$ 125 & 1.94	$\pm$ 0.027 &	1.40	&		5.01 $\pm$	0.16\\
		5871 & 11006& 6500 $\pm$ 125 & 1.39 $\pm$ 0.019 &	1.40       &		2.98 $\pm$	0.09\\
		5426 & 17028& 6250 $\pm$ 125 & 1.08 $\pm$ 0.015 &	 & 	1.46 $\pm$	0.04 \\
		\hline
	\end{tabular}
\end{table*} 

\paragraph*{}

\begin{figure*}
 	\centering
 	\includegraphics[width= 2.2\columnwidth]{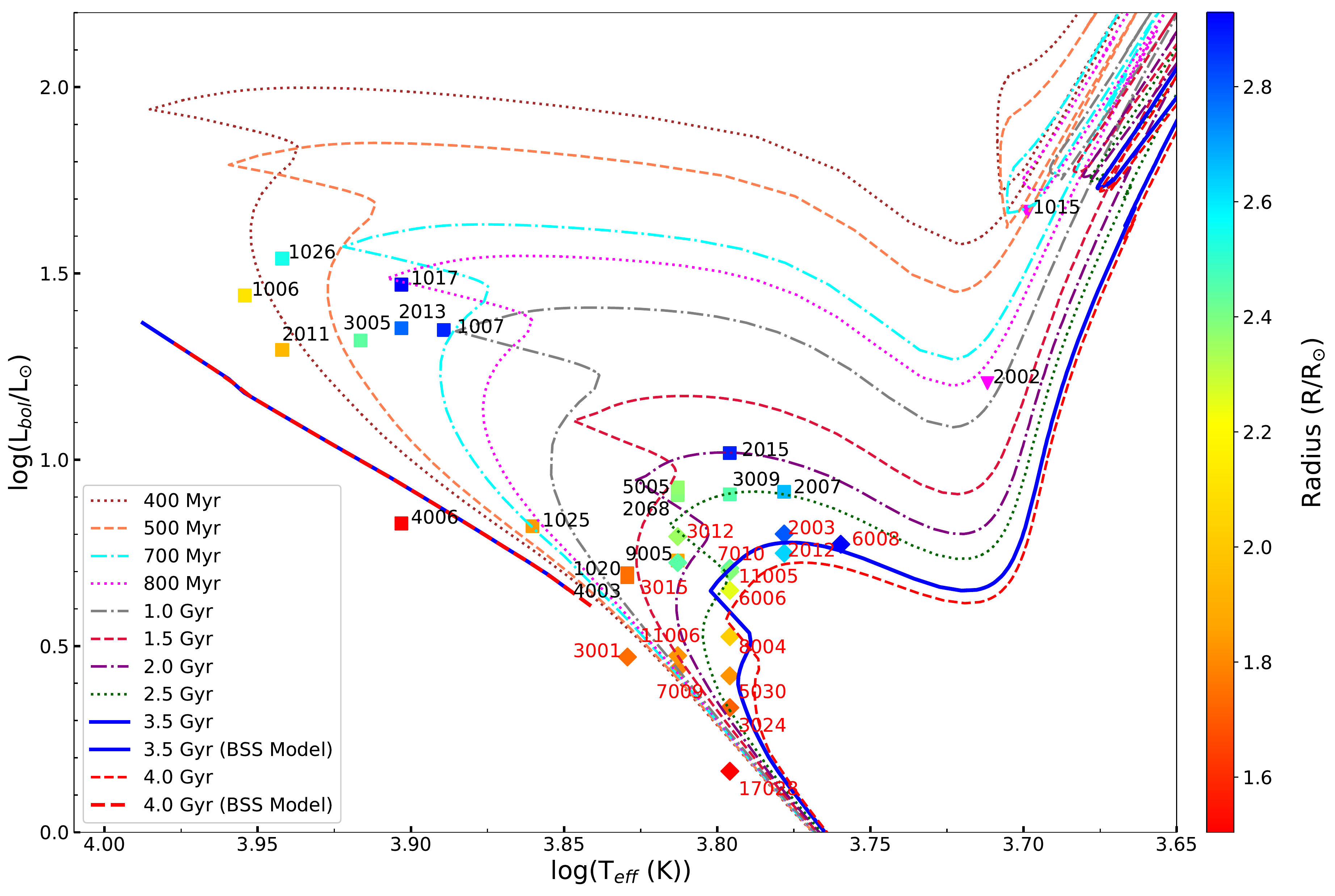}
 	\caption{Luminosity vs. Effective temperature, H-R diagram of the BSSs and FUV bright stars. The points are colour coded based on the radius of the stars, shown on the right. 2 YSSs (pink triangles) are also shown in the plot. The BSSs are labelled in black and FUV bright stars in red. Isochrones generated using FSPS for various ages are shown with different colours. The BSS model line for 3.5 and 4 Gyr are also shown in the figure.}
 	\label{LvsT for BSS}
 \end{figure*}

We used the Hertzsprung - Russel Diagram (L/L$_{\sun}$ vs T$_{eff}$) and L/L$_{\sun}$vs R/R$_{\sun}$ diagram to compare the properties of the BSSs and the FUV bright stars. In Figure \ref{LvsT for BSS}, the L/L$_{\sun}$ vs T$_{eff}$ plot is shown, where the BSSs, 2 YSSs and the FUV bright stars are shown with different symbols along with their identification numbers. These stars are also colour coded according to their radius. The BSSs and YSSs are labelled in black and the FUV bright stars are labelled in red. We calculated L/L$_{\sun}$ of 2 YSS with the values of R/R$_{\sun}$ and T$_{eff}$ taken from \citet{Leiner2016} and \citet{Landsman1997_S1040}. In this figure, we over plotted the isochrones of various ages, and BSS model lines for 3.5 Gyr and 4 Gyr population of BaSTI model generated by FSPS. 

In Figure \ref{LvsT for BSS}, we notice that the 7 BSSs are indeed quite luminous and are well separated from the other BSSs. All these are bright in the FUV and are the brightest 7 among the BSSs. The rest of the 10 BSSs are located between the BSS model line and the isochrone, with relatively low luminosity. We notice that all the BSSs are either hotter or luminous than the 3.5 Gyr isochrone. We can possibly group the BSSs into 3 groups: (a) the luminous and hot BSSs (WOCS (1006, 1007, 1017, 1026, 2011, 2013 and 3005)), (b) hot and less luminous BSSs (WOCS (4006, 4003, 1025,1020 and 9005)) and (c) moderately luminous and cooler BSSs (WOCS (3009, 5005, 2068, 2015 and 2007)). The group (a) BSSs are found to be located in the isochrones with the age range of 400 Myr - 1 Gyr. Their younger age is consistent with their higher temperature, luminosity and mass. This probably is why group (a) BSSs are FUV bright. Thus the group (a) BSSs are likely to be youngest among the BSS population. WOCS 1007 is a group (a) BSS with a relatively large radius suggestive of it being more evolved. This is a $\delta$-Scuti star (\cite{Gilliland1992}) and they suggested that this star could be evolving from the MS to the subgiant phase. The YSS (WOCS 1015) is found to be located near the isochrone of the range 500 Myr - 700 Myr. \citet{Leiner2016} estimated the age of a YSS  (WOCS 1015) to be in the range of 450 -$\sim$ 750 Myr, consistent with our estimation. We also find that WOCS 2002 has an age of 800 Myr - 1 Gyr. \citet{Landsman1997_S1040} suggested a possible evolutionary sequence for this binary, with $\sim$ 765 Myr as the period of stable mass transfer for the formation of the BSS progenitor, and a further 75 Myr for its post-BSS evolution. This is consistent with our age estimation. We suggest that the two YSSs could also belong to the group (a) BSSs.

 \begin{figure*}
	\centering
	\includegraphics[width= 2.2\columnwidth]{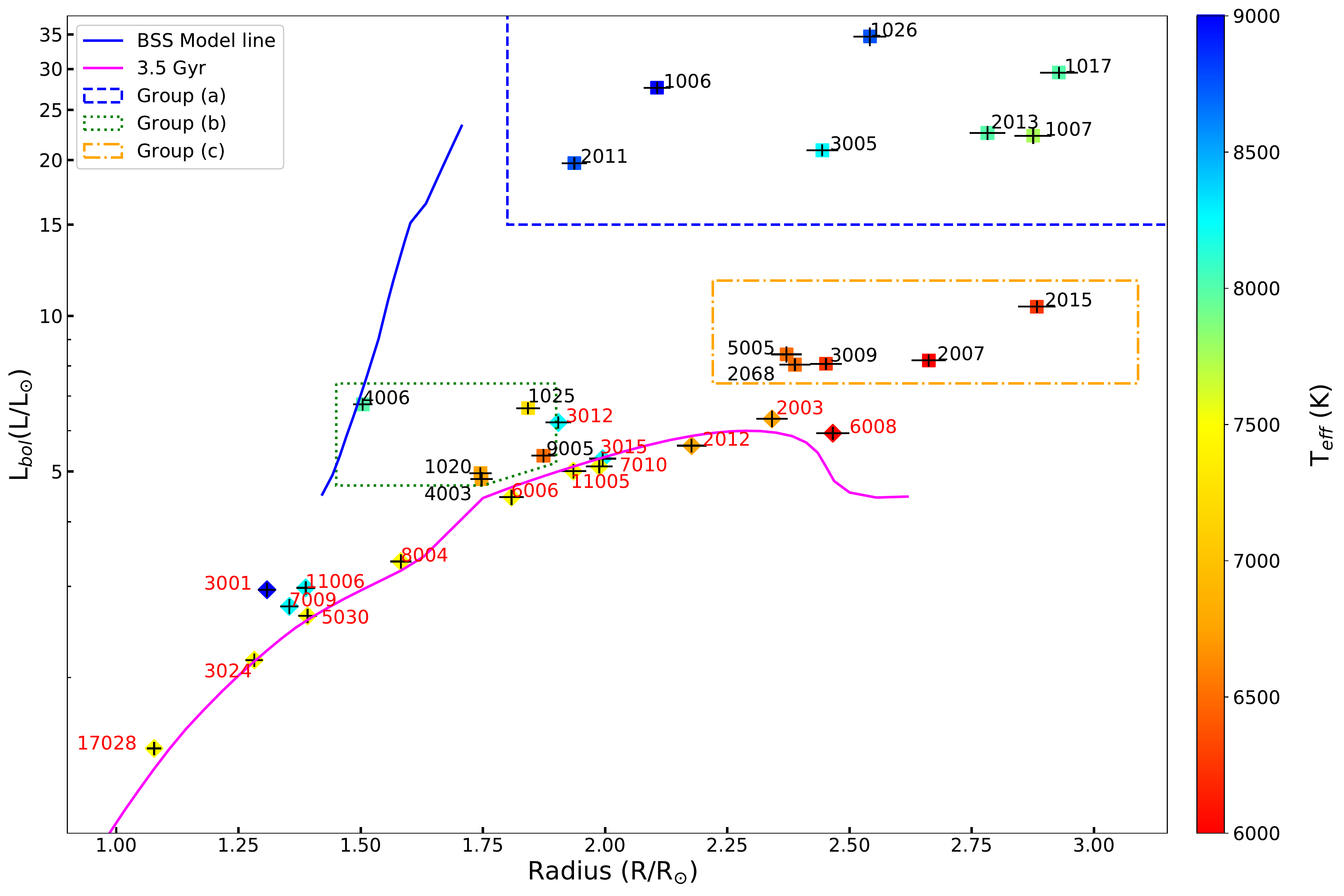}
	
	\caption{A plot of Luminosity vs. Radius for BSS and FUV bright stars. The stars are colour coded based on the effective temperature shown on the right side of the figure. The BSS are labelled in black and FUV bright stars in red.}
	\label{LvsR plots for BSS}
\end{figure*}

\begin{figure*}
	\centering
	\includegraphics[width= 1.8\columnwidth]{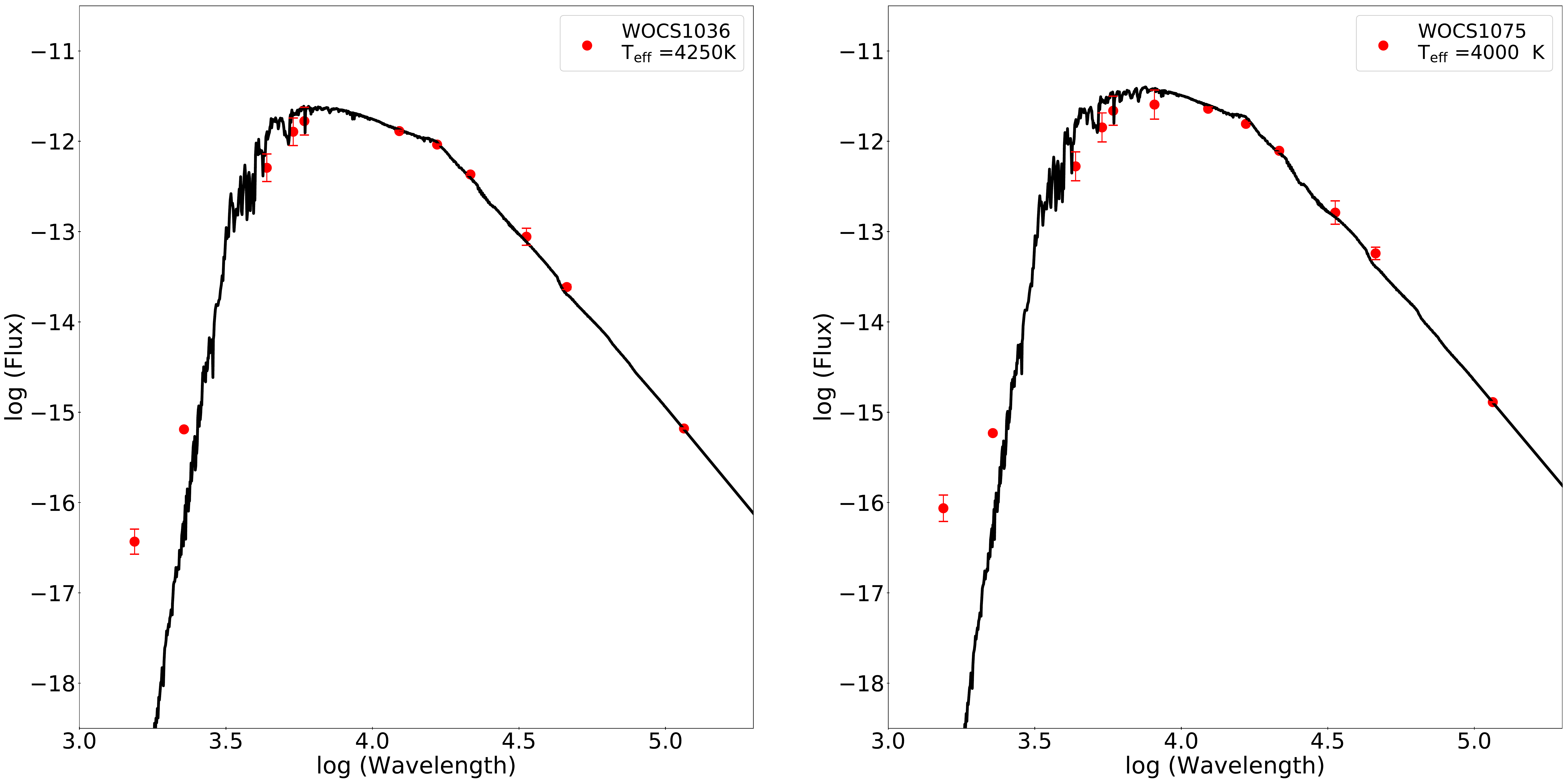}
	\includegraphics[width= 1.85\columnwidth]{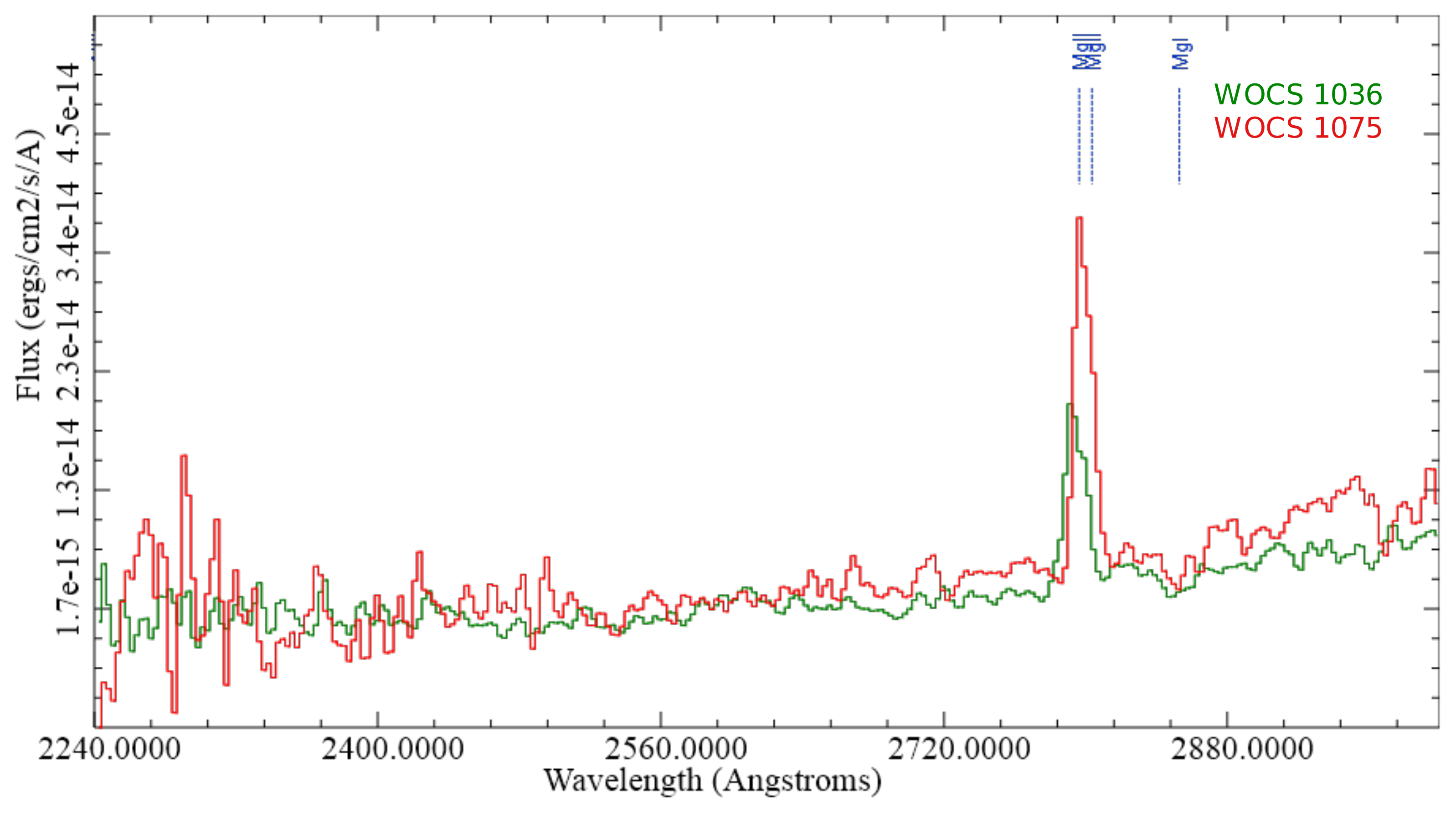}
	\caption{ (Top) SEDs (extinction corrected) of RGs (WOCS 1036 and WOCS 1075 - detected in the FUV band) with photometric flux from UV to IR. The fitted spectra suggest that there could be excess in both NUV and FUV for these two RGs. Best fitting Kurucz model spectrum is overplotted and the corresponding temperature is listed in each panel. The unit of wavelength is \angstrom\, and flux is \flux. (Bottom) IUE spectra of these RGs  in the NUV region having the Mg II {\it h} + {\it  k}  lines in emission.}
	
	\label{fig:SED of RG}
\end{figure*}

\begin{figure*}
	\centering
	\includegraphics[width=2.0 \columnwidth]{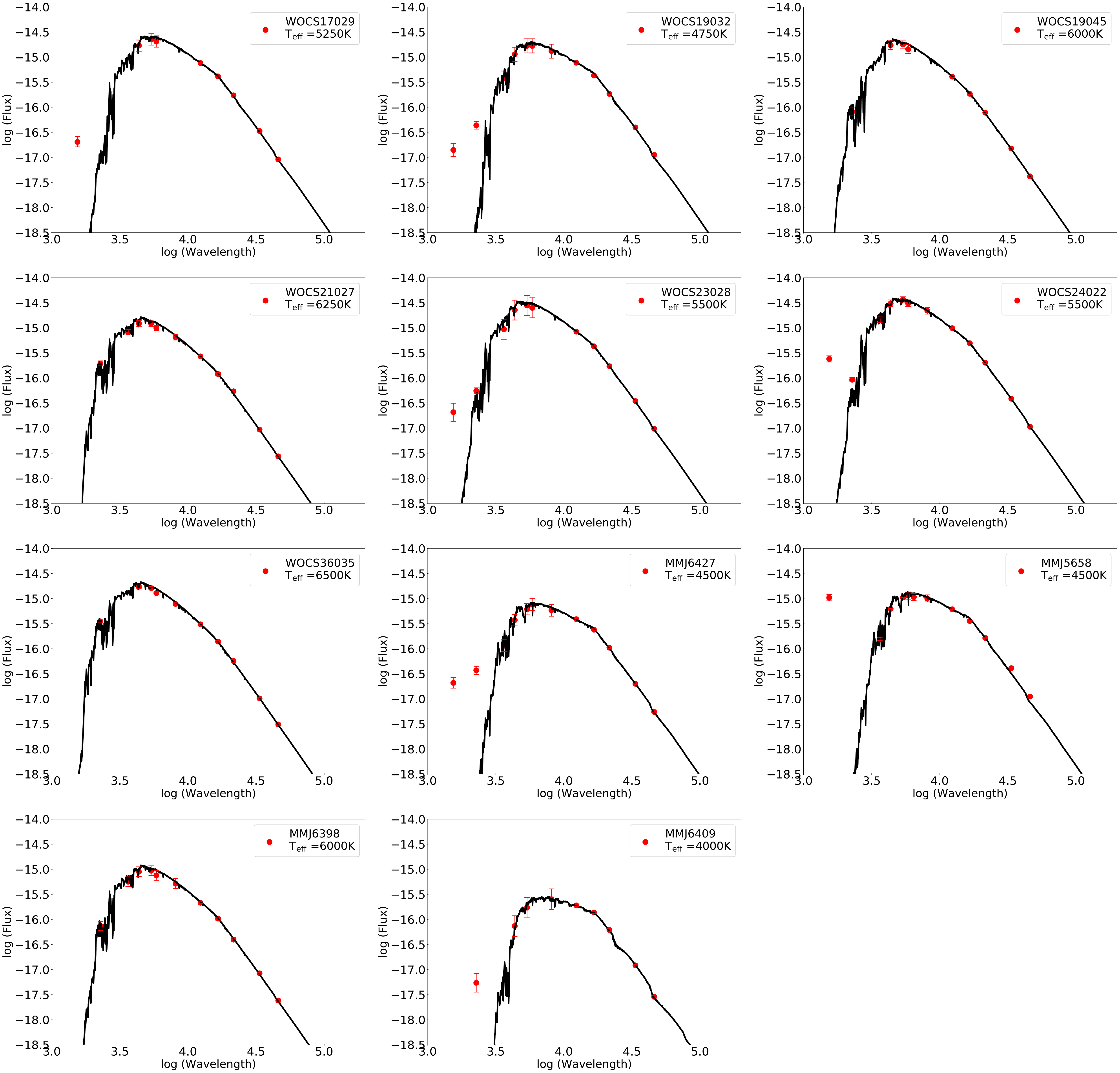}
	\caption{SEDs (extinction corrected) of WD+MS candidates with photometric flux from UV to IR. Best fitting Kurucz model spectrum is overplotted and the corresponding temperature is listed in each panel. The unit of wavelength is \angstrom\, and flux is \flux.}
		\label{fig: SED of WD+MS}
\end{figure*}
 \begin{figure*}
	\centering
	\includegraphics[width=1.8\columnwidth]{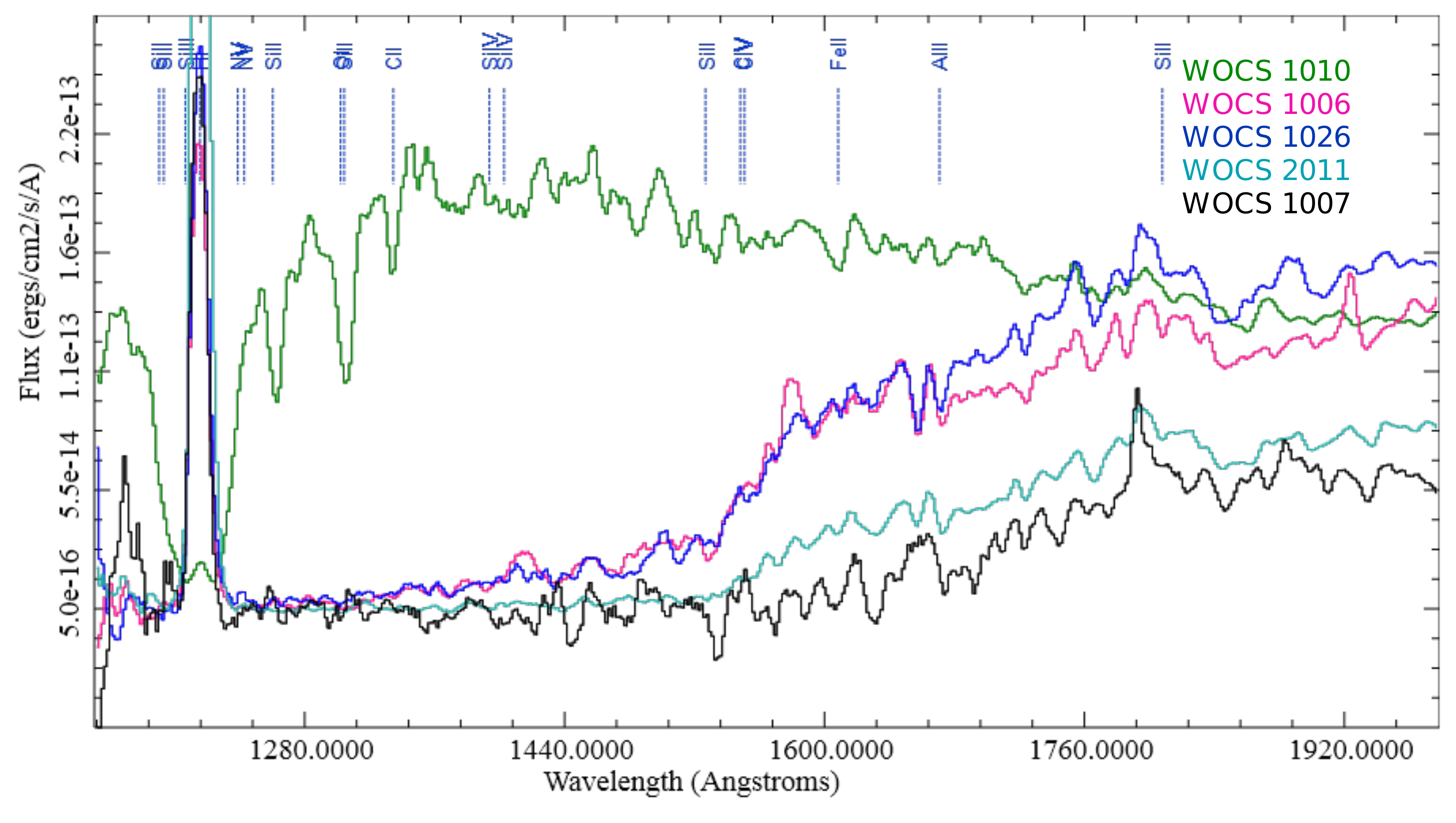}\\
	\includegraphics[width=1.8\columnwidth]{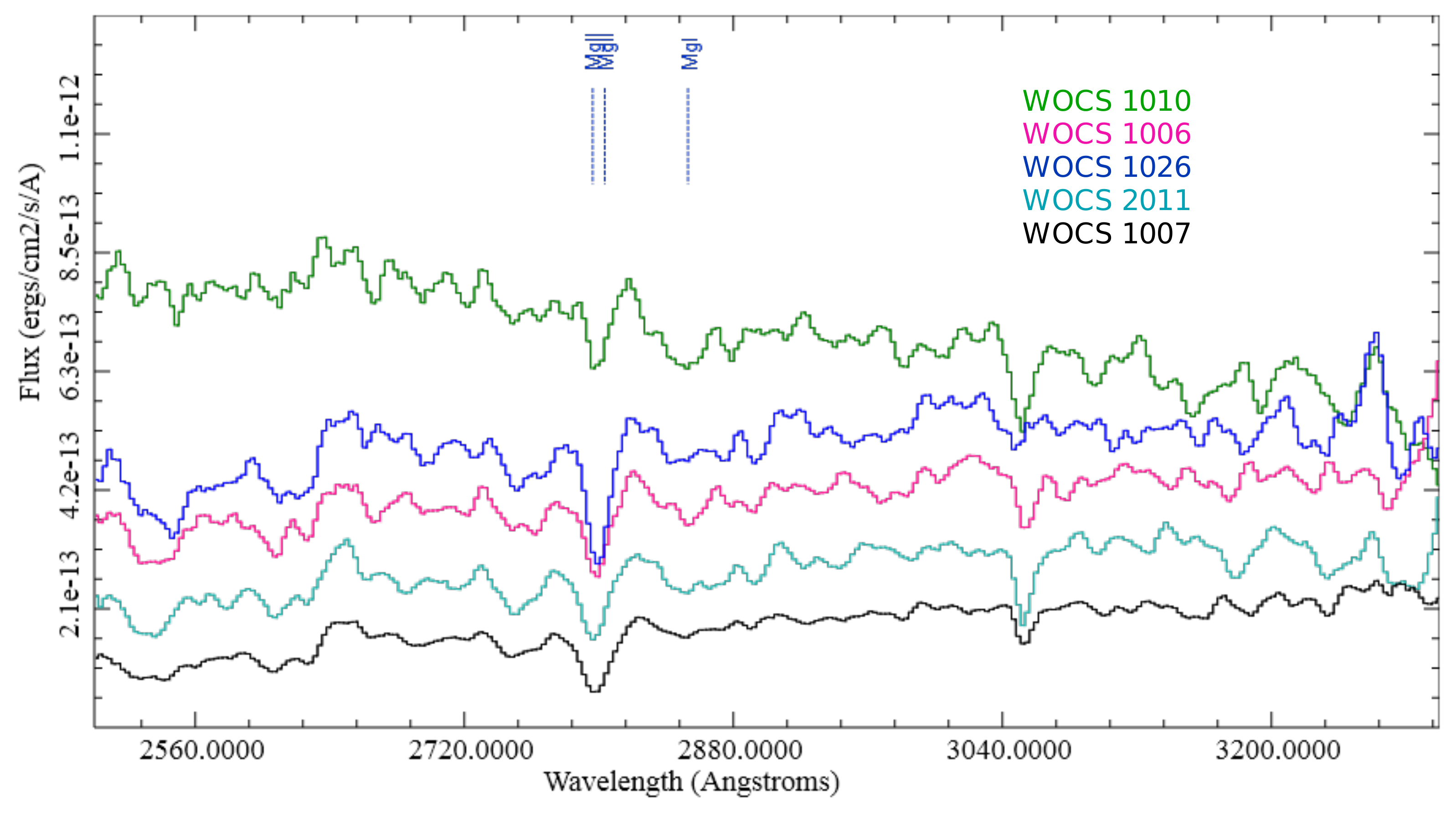}
	\caption{Upper panel: IUE spectra in the FUV for 5 BSS. The WOCS 1010 spectrum is scaled down by 0.1 to plot along with the other spectra. The WOCS 2011 is scaled up by 1.5 as well. The spectra of WOCS 1006, WOCS 1007  and WOCS 1026 are shown without any scaling. Some of the spectral lines are marked. Lower panel: The spectra of 5 BSS in the NUV range. The spectra are scaled to bring out the absorption profile of the MgII {\it h} +{\it k} line, which is present in all the spectra.}
		\label{fig: allbss_IUE_FUV}
\end{figure*}
Among the group (b) BSS, WOCS 4006 is known to be a $\delta$-scuti star (\cite{Gilliland1992}). We estimate the star to be slightly under luminous, with respect to the previous estimates. The remaining stars in the group are located in and around the plotted isochrones, with moderate radii. As these stars are located close to the MS and as well as BSS model line, we are unable to estimate their age range. This suggests that the group (b) BSSs do not show any evidence of evolution beyond the MS. The group (c) BSSs, on the other hand, are located on the subgiant branch of the plotted isochrones with an age range 2-2.5 Gyr. These BSSs are relatively cooler and have larger radii, which are consistent with their location on the isochrones. Thus, group (c) may be an evolved group of old BSSs with an age range of 2-2.5 Gyr.
  
  We notice that all of the FUV bright stars are located near the older isochrones. Among them, WOCS 3001, WOCS 11006, WOCS 7009 and WOCS 17028 are located close to the MS of the isochrones, though WOCS 3001 and WOCS 17028 are marginally under-luminous. We could consider them to be similar to the group (b) BSSs. Therefore, there is a possibility that the FUV bright stars could actually be BSS like stars. In Figure \ref{LvsR plots for BSS}, we present L/L$_{\sun}$ vs R/R$_{\sun}$ plot for the BSSs and the FUV bright stars, where the stars are colour coded based on T$_{eff}$. We have marked the regions occupied by these three groups in the figure. In this figure also, the group (a) BSSs are located separately, whereas the group (b) and (c) show a similar trend, with group (c) stars located towards the right extreme. It is interesting to note that the FUV bright stars, and group (b) and (c) stars have similar location in this diagram. The $\delta$-scuti star (WOCS 4006) is found to be located away from the sequence, suggesting that it probably has lesser radius (as it is already found to be under luminous). From this figure, we can only infer that the following FUV bright stars share the same location as the BSSs; WOCS (6008, 2003, 2012, 3012, 3015, 11005, 6006 and 7010).

We also created SEDs for the two RGs which show FUV excess, which are shown in the upper panels of Figure \ref{fig:SED of RG}. Both stars show excess flux in the NUV and FUV, with respect to the fitted spectrum. This is suggestive of the presence of excess UV flux in these RGs, which explains them having detected in the FUV band. The SED fit suggests a temperature of  $\sim$ 4000K for the RGs. The FUV flux detection is close to the detection limit of GALEX. The origin of the UV excess flux could be due to chromospheric activity or the presence of a hotter component. IUE observations of some of the M67 RGs were done to detect and characterise their chromospheric emissions. An archival search finds that both these RGs have IUE spectra. WOCS 1036 (NGC2682 4202/ M67 IV-202) was observed on 28/10/1984, in the NUV region for an exposure time of 24.6 ks. WOCS 1075 (NGC2682 S1553) was observed on 14/03/1987, in the NUV region for an exposure time of 14.7 ks. These two spectra are shown in Figure \ref{fig:SED of RG}. Both the spectra show the presence of Mg II {\it h} + {\it  k} lines in emission, which are marked in the figure. As these are low resolution spectra, the two Mg II {\it h} + {\it  k} lines are not separated. IUE spectra of 5 more RGs are also available and an inspection of their Mg II {\it h} + {\it  k} line suggests that mild emission is detected in WOCS 1008 and WOCS 1045, whereas no emission is found in WOCS 1005, WOCS 1054 and WOCS 2059. The presence of Mg II {\it h} + {\it  k} lines in emission is suggestive of chromospheric activity. \citet{2011_Martinez} studied 177 cool G, K and M giants and supergiants using IUE spectra and measured their Mg II {\it h} + {\it  k} line strength. They argue that these emission represent the chromospheric radiative energy losses presumably related to basal heating by the dissipation of acoustic waves, plus a highly variable contribution due to magnetic activity. The detection of the Mg II {\it h} + {\it  k} emission lines along with the presence of excess continuum UV flux suggests that the FUV detected RGs have more chromospheric activity, in comparison to the other RGs in the cluster. 
	
	We also constructed SEDs for the MS+WD candidates (Figure  \ref{fig: SED of WD+MS}). We detect excess flux in the UV for 7 of the 11 stars. We observe a range in the excess FUV flux detected in these stars. Thus the UV excess found in 7, suggests that, these could be potential MS+WD binaries. The rest of the 4 sources do not have FUV flux and are also not found to have any excess in the NUV flux, with respect to the fitted SED. The slope of the flux in the UV, is suggestive of a range in temperature for the WD companions. The temperature of the MS stars are found to be in the range 4000K - 6250K. The 4 stars which are not found to have any UV excess are located to the left of the MS in the optical CMD (Figure \ref{optical CMD}). It is interesting to note that the estimated temperatures of these stars are at least 1000 K higher than the MS stars at the same luminosity. The fact that these stars are bluer with respect to the MS in optical and NUV$-$V CMDs due to their inherent hotter temperature might warrant a closer look at the nature of these stars.
	
\section{Discussion}
\label{discussion}
M67 is a very well studied cluster and these studies have revealed the presence a large number of stars in non-standard stellar evolutionary phases. Some of them being, pulsating BSSs, YSSs, SSGs, W Uma systems, triple systems etc. Many single stars are also found to be PVs or RRs. Therefore M67 serves as a test bed to study these stars in detail and understand their properties, origin and evolution. It is thus essential that the properties of these systems be studied in a broad wavelength range to check whether they comply with properties estimated mainly from the optical pass bands. There have been extensive studies of this cluster in the optical and  X-ray, but a detailed study in the UV pass bands is not attempted so far. In this study, we use the deep photometric data obtained from GALEX in the NUV and the FUV pass bands to study the behaviour of M67 member stars. Here, we present the UV CMDs and SEDs of some selected stars.

We combined all the observations of M67 made by GALEX to obtain the FUV and NUV magnitudes of 449 stars. We used the optical CMD as the reference to deduce the evolutionary status of stars and compare their location in the UV CMDs to understand their behaviour. We detect 92 stars in the FUV, which include BSSs and WDs.  Most of the FUV detected BSSs are bright and located 3 - 8 magnitudes above the MSTO in the UV CMD. As the BSS region in the UV CMDs extends to about 8 mag above the MSTO, these CMDs are ideal to study BSSs. This was earlier noticed by \citet{Sindhu2015} and \citet{seigel14}. The CMDs shown in figure \ref{FUV-optical CMD}, is a first of its kind for this cluster and it reveals a lot of interesting properties of various types of stars. The sequences which are identified in the UV CMDs suggest that there are a good number of stars with FUV and/or NUV excess. 

 We empirically idenitifed regions in the FUV$-$V CMD belonging to WDs, gap stars and BSS. In the BSS region, it was noticed that 15 stars were co-located along with the BSS. The BSS are in general identified in the optical CMD as stars located above the MS. Now, if we apply the same criteria in the UV CMD, we find a lot more stars in the BSS region. There may be FUV/NUV non-photospheric emission resulting in a UV excess occuring in these stars appearing brighter than the MSTO.  Also, it is better to identify BSSs based on their total luminosity, rather than their brightness in a particular pass band. Thus, we make use of the H-R diagram to understand the properties of BSS as well as FUV bright stars. We detect 15 stars to have similar UV flux as the BSSs. We estimated L/L$_{\sun}$, R/R$_{\sun}$ and T$_{eff}$ for the GALEX detected BSSs and the FUV bright stars using SEDs. The figures \ref{LvsT for BSS} and \ref{LvsR plots for BSS} suggest that the BSSs in M67 have a wide range of properties. We use these figures to group the BSSs into 3. The location of BSSs in the H-R diagram is interpreted with the help of overlaid isochrones. The BSSs are found to be located in isochrones of a large age range (400 Myr to 2.5 Gyr). The group (a) BSSs are found to be hotter, luminous and the youngest with an age range of 400 Myr - 1 Gyr, whereas the group (c) stars are as old as 2.5 Gyr. The group (a) BSS are found to be brighter up to 10 magnitudes in the FUV. We suggest that the large range of FUV magnitude of BSSs is likely due to their range in their temperature and luminosity. As we find the BSSs to be more or less evenly distributed across the large age range, the BSSs is M67 is likely to be forming fairly continuously in the 400 Myr - 2.5 Gyr duration. 
	
The figure \ref{LvsR plots for BSS} gives a different perspective of the BSS, with respect to the expected location of BSS. The group (b) and (c) mostly follow the 3.5 Gyr isochrone values for their L/L$_{\sun}$, R/R$_{\sun}$, with only one star following the BSS model line. All the group (a) BSSs are found to have relatively larger radii. If we consider only the 6 FUV bright stars which lie very close to the BSS model line in the FUV$-$V CMD (See figure \ref{FUV-optical CMD}) to check whether any of these could be BSS, we see that, these stars have T$_{eff}$ very similar to the co-located BSS. Among these WOCS 3012 lies within the group (b) location, and is a potential BSS candidate. The star WOCS 6008 has a large radii and is located below the group (c) location, whereas WOCS 3001 and WOCS 11006 are found to be hotter than the MS, suggesting that these 3 stars could also be BSS candidates. On the other hand, WOCS 11005 and WOCS 2012 are located on the MS and hence are unlikely to be BSS. Thus, 4 among the FUV bright stars could be BSS candidates. The fainter stars in the FUV$-$V CMD, located in the BSS region could be FUV bright due to a variety of reasons. We also detect YSS stars, wheres the RGs are located just outside this region. It is likely that this group might comprise of a wide variety of stars which show UV excess due to non-photospheric emission, as well as BSS.

 The BSSs in M67 spans a very large range in T$_{eff}$, unlike those in NGC 188 (\citealp{gosnell15}). This is one of the reasons for a large range in their FUV magnitude. The SEDs of a few of the BSSs including WOCS 2015 have FUV excess and it is important to understand the origin of this excess. Some of the BSSs in NGC 188 is identified to have a hot companion, resulting in FUV excess (\citealp{gosnell15}, \citealp{2016Subramaniam}). Thus, it will be interesting to probe whether these BSSs also have any hot companion. We searched the IUE archive for spectra of bright BSSs and we could trace low resolution spectra of 4 BSSs. These 4 BSSs (WOCS (1006, 1007, 1026 and 2011)) along with the UV bright BSSs WOCS 1010 have FUV as well as NUV spectra. The FUV spectra of these 5 BSSs are shown in Figure \ref{fig: allbss_IUE_FUV} (top panel). The FUV spectrum of WOCS 1010 is scaled down by 0.1 to place it along with the other spectra, which is suggestive of its large UV flux. This spectrum has very well defined absorption lines and is also suggestive of a relatively hotter temperature relative to the other BSSs. The spectra of BSS WOCS 1006, WOCS 1007 and WOCS 1026 appear very similar with respect to the continuum as well as absorption lines, and all of which belong to group (a). Some absorption lines can be identified in all the spectra. The spectrum of WOCS 2011 has relatively low signal and is scaled up by 1.5. This spectrum does not show any clear absorption lines. In the bottom panel, the NUV spectra are shown. All the 5 BSSs show the presence of Mg II {\it h} +{\it k} line in absorption. This is a clear indicator of absence or very low chromospheric activity for these BSSs. Thus, if there is any excess in the UV emission present in these BSSs, it may not be due to chromospheric activity, unlike the RGs.
 
 We also find some of the FUV bright stars to have excess UV flux. These could harbour hot companions or have active chromospheres. The SED analysis of the MS+WD candidates also suggest hot companion. As the GALEX FUV gives one data point in the FUV region of the SED, it is hard to characterise the hot component based on a single data point, even though the excess is much larger in a few cases. \citet{2016Subramaniam} demonstrated that UVIT observations with its filters in the FUV and NUV can provide a good profile of the UV continuum of stars with UV excess. We plan to obtain observations in the multiple filters of UVIT. This will help in detecting the WDs if they are present.  These observations will surely help in placing limits on the flux contribution in the FUV due to chromospheric activity and the temperature range of chromospheres of stars in M67.
 
 Detection of a large number of stars with FUV and/or NUV excess in this cluster suggests that these stars could be chromospherically active. We detect 2 RGs with FUV excess in concurrence with the presence of Mg II {\it h} +{\it k} emission lines from the archival IUE spectra, suggestive of significant chromospheric activity. Recently, \citet{Stello2016} found these two giants to have very low frequency oscillations, whereas some other giants in this cluster were found to be oscillating with relatively large frequency. Connecting this to the dispersion in the location of RGs in the (NUV$-$V) CMD, we speculate that some RGs in this cluster are chromospherically active resulting in excess flux in the NUV. The physical processes operating in the cool giants with active chromospheres are still being understood (\citealp{2011_Martinez}). The excess emission seen could be due to the dissipation of purely mechanical energy in the turbulent chromosphere or magnetic activity. As one expects minimal magnetic activity in the giants, the heating and emission from the chromosphere of these red giants can be mostly attributed to the dissipation of mechanical energy.  It may be possible that the low frequency waves seen in the active red giants, effectively contribute to the heating of chromosphere. A discussion on possible modes of oscillations and energy dissipation rate in red giants can be found in \citet{Dziembowski2001}.
 
 A number of stars near and below the MSTO are found to have NUV excess flux. All MS stars in the F-M spectral types are observed to have varying amounts of chromospheric activity (\citealp{Linsky2017}). \citet{Reiners2009} studied 15 solar-type stars and concluded that high activity in these stars that exceeds the solar levels is likely to be due to rapid rotation. This study thus clearly brings out the need to probe cluster stars in the UV wavelength. 

\section{Conclusions}
\label{Conclusions}
The main conclusions of this study can be summarised as follows:
\begin{enumerate}
	\item{The first comprehensive UV properties of member stars of the old open cluster M67 are presented here. We have compiled the UV magnitudes of 449 stars (92 in the FUV and 424 in the NUV). We have provided an online catalog with FUV and NUV magnitudes, and classification of the detected member stars.}
	\item{16  BSSs, 3 WDs, 3 YSSs, one SSG, two RGs, 3 triple system in the FUV and 13  BSSs, 2 WDs, 2 YSSs, 2 SSGs and 1 triple system in the NUV.}
		\item {We detect a few lower MS stars to have large UV excess, suggesting that they could be MS+WD binaries. We identified 11 such systems, which are new identifications, with 7 showing FUV excess.}
		\item{ We also detect stars near the MSTO and subgiant branch to have excess flux in the FUV and/or NUV.  This excess flux could be due to chromospheric activity. 15 stars are found to have relatively large excess in the FUV, which are found to be as bright as the BSSs in the FUV.}
	\item {The BSS in M67 span a large range in T$_{eff}$ and L/L$_{\sun}$, which is found to correlate well with the large range of FUV magnitude (14-23 mag). The FUV bright BSSs are found to be the hotter, luminous, and probably younger BSSs.}
	\item{The isochrones overlaid on the H-R diagram suggest that the group (a) including the two YSSs is very young (400 Myr - 1 Gyr). WOCS 2011 and WOCS 1006 could be the most recently formed BSS,  good candidates to probe the BSS formation mechanism. The group (c) stars are likely to be the oldest among the BSS in M67. We suggest that the BSS in M67 are formed in the last 2.5 Gyr - 400 Myr, more or less continuously.}
	\item{We detect 2 RGs to have UV continuum excess as well as emission in the Mg II {\it h} +{\it k} lines from the IUE archival spectra. Along with the scatter of the RG stars in the NUV$-$V CMD, and the detection of stars with UV excess, we speculate that the RGs as well as a good fraction of stars in M67 could be chromospherically active. The bright BSSs stars are not found to be chromospherically active.}
\end{enumerate}

 \section*{Acknowledgements}
The authors like to thank Prof. Ram Sagar for the useful discussion. This research has made extensive use of Vizer and VOSA. Sindhu N acknowledges support from CSIR for the SRF fellowship through grant  3/890(0005)/17 EMR-I. We thank the anonymous referee for the insightful comments which greatly improved the manuscript.


\bibliographystyle{mnras}
\bibliography{references} 






\bsp	
\label{lastpage}
\end{document}